\documentclass[1p]{elsarticle}

\usepackage{lineno,hyperref}
\modulolinenumbers[5]

\journal{Journal of Geometry and Physics}

\usepackage{graphicx}
\usepackage{amsmath}
\usepackage{amsthm}
\usepackage{appendix}
\usepackage{algorithm}
\usepackage{algpseudocode}
\usepackage{booktabs}

\theoremstyle{definition}
\newtheorem{definition}{Definition}

\begin{document}

\begin{frontmatter}

\title{On the curvatures of Gaussian random field manifolds}

\author{Alexandre L. M. Levada}
\address{Computing Department, Federal University of S\~ao Carlos, S\~ao Carlos, SP, Brazil}
\ead{alexandre.levada@ufscar.br}

\begin{abstract}
Information geometry is concerned with the application of differential geometry concepts in the study of the parametric spaces of statistical models. When the random variables are independent and identically distributed, the underlying parametric space exhibit constant curvature, which makes the geometry hyperbolic (negative) or spherical (positive). In this paper, we derive closed-form expressions for the components of the first and second fundamental forms regarding pairwise isotropic Gaussian-Markov random field manifolds, allowing the computation of the Gaussian, mean and principal curvatures. Computational simulations using Markov Chain Monte Carlo dynamics indicate that a change in the sign of the Gaussian curvature is related to the emergence of phase transitions in the field. Moreover, the curvatures are highly asymmetrical for positive and negative displacements in the inverse temperature parameter, suggesting the existence of irreversible geometric properties in the parametric space along the dynamics (\emph{the curvature effect}). Furthermore, these asymmetric changes in the curvature of the space induces an intrinsic notion of time in the evolution of the random field.
\end{abstract}


\end{frontmatter}


\section{Introduction}

In science, many natural phenomena are modeled by a stochastic complex system in which collective properties emerge from the non-linear interactions among its parts in different levels of scale \cite{Paolo}. Typically, in these systems, the whole is more than the sum of the parts. In this context, the characterization of the dynamics of stochastic systems using a mathematical tool is crucial for understanding the underlying processes that govern the emergence of complex behavior \cite{Baryam}. For example, being able to predict when a phase transition is going to happen is a relevant and challenging problem in complex systems analysis.

There are several mathematical models for the study of complex systems, as cellular automata \cite{Automata}, complex networks \cite{ComplexNetworks} and random fields \cite{RandomFields}. Particularly, random field models are employed for studying non-deterministic phenomena in which non-linear interactions between random variables lead to the emergence of long range correlations and phase transitions \cite{Willsky,Phase}. Random fields arise naturally in several areas of science, such as statistical mechanics \cite{StatMech}, thermodynamics \cite{Gibbs}, biology \cite{MRFBiol} and economics \cite{MRFEcon}.

Several random field models consider that the random variables can assume a finite and discrete number of states, such as the Ising \cite{Ising} and the $q$-state Potts model \cite{Potts}. In this paper, our focus is in the study of Gaussian random fields, where each variable can assume any value belonging the real line, that is, the set of possible states is infinite and continuous \cite{GaussianRandomFields}. The main objective of this scientific investigation is to propose an information-geometric framework to understand and characterize the dynamics of Gaussian random fields defined on two-dimensional lattices.

In particular, we assume some simplifying hypothesis: first, the random field is Markovian in the sense that the probability of a given variable in the field depends only on the variables belonging to a local neighborhood system around the variable \cite{MRFApp}. Second, the model is isotropic in the sense that the inverse temperature parameter, which control the spatial dependence structure, is spatially invariant and constant for all orientations in space. Last, but not least, we deal with a pairwise interaction model, which means that we allow only binary relationships. In summary, we consider a pairwise isotropic Gaussian-Markov random field to model the interaction between spatially dependent Gaussian random variables organized in a 2D lattice. The degree of interaction is quantified by a coupling parameter, also known as the inverse temperature. It has been observed that an increase in the inverse temperature parameter beyond a critical value induces the emergence of phase transitions in the system.

In the analysis of complex systems, one question that naturally arises is: how can we predict that the system is approaching a phase transition in a quantitative way? Besides, how to determine whether the system is moving towards order or randomness? Often, researchers extract several statistical and physical measures such as entropy, free energy, average magnetization, etc. from both systems and use an extrinsic distance function (Euclidean distance) to measure a degree of similarity between the feature vectors. The proposed method allows the computation of intrinsic geometric properties from the parametric space, such as the first and second fundamental forms (Fisher information matrices) of the underlying manifold.

During the 1950's, John Wheeler was a pioneering theoretical physicist in geometrodynamics, a research field whose main goal is to characterize and describe spacetime and related phenomena completely in terms of geometry \cite{Wheeler}. His attempt was to unify the fundamental forces and reformulate general relativity as a configuration space of three-metrics. These issues have been investigated by several physicists and remain an active field in the 21st century, as a mathematical tool for the unification of gravitation \cite{Geom2014}, quantum mechanics \cite{Unification} and in the study if particle systems \cite{Nature}. In this context, our study can be thought as an attempt to describe the dynamics of Gaussian random fields completely in terms of intrinsic geometric properties of their parametric spaces. Another motivation for the proposed methodology is information geometry, a research field that combines information theory and differential geometry to study intrinsic geometric properties of parametric spaces associated with random variables \cite{Amari,Frieden2004,Dodson}. The main goal of this paper is to build computational tools in order to study how the Gaussian, mean and principal curvatures change as the inverse temperature parameter increase. At the beginning of the dynamics, the inverse temperature parameter is zero, and the random field model degenerates to a regular Gaussian distribution, once we have a collection of independent random variables. In this scenario, the parametric space exhibit constant negative curvature (hyperbolic geometry) \cite{Pinele}, which means that the first and second fundamental forms are identical. However, when the inverse temperature increases, the underlying parametric space suffers severe geometric deformations. Little is known about this process in situations where the random variables are not independent. The idea is to analyze how the emergence of a spatial dependence structure along time leads to geometric transformations in the parametric space. The obtained results show that the variations of the Gaussian curvature when the system moves towards higher entropy states is different from the variations of the Gaussian curvature when the system moves towards lower entropy states, which induces an intrinsic notion of time as a natural one-way direction of evolution. In other words, to decrease the system's entropy, the parametric space must be stretched/shrunk harder when it is bending.

The remaining of the paper is organized as follows: in Section 2, we define Gaussian random fields, showing that it is a curved statistical model. In Section 3, we present important differential geometry concepts, such as surface, tangent planes, fundamental forms, Gaussian, mean and principal curvatures. Section 4 defines the first and second-order Fisher information matrices, their relation with the fundamental forms and presents the complete derivation of the components of these operators in pairwise isotropic Gaussian-Markov random fields. Section 5 shows the results obtained by computational experiments using MCMC simulations. Finally, Section 6 presents the conclusions and final remarks.

\section{Gaussian random fields}

Random fields are mathematical models used in the study of stochastic and non-linear complex systems \cite{RFields}. Among all the random field models categorized in the literature, Gaussian random fields are remarkably important \cite{GRFs}. Part of this relevance is due to the central limit theorem \cite{CLT}, which states that the summation of independent and identically distributed random variables tend to be normally distributed. Several random field applications can be found in the literature, from image processing techniques in computer vision \cite{MRF_DNN} to statistical and cosmological physics \cite{Review_Ising,Peacock}.

\begin{figure}[ht]
	\begin{center}
	\includegraphics[scale=0.4]{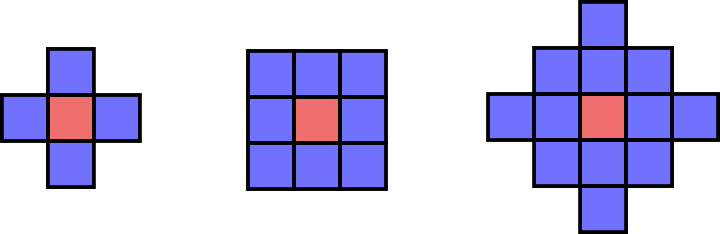}
	\end{center}
	\caption{First, second and third order neighborhood systems on a 2D lattice.}
	\label{fig:neigh}
\end{figure}

Pairwise isotropic Gaussian-Markov random fields (GMRF's) are mathematical structures particularly suitable to study spatially dependent continuous random variables by means of non-linear interactions between neighboring particles on a lattice. The main advantage of this model is related to the mathematical tractability. First, in this model, all the parameters are scalars, making the parametric space a regular manifold. Moreover, we can derive closed-form expressions for several expected values, which makes it possible the exact computation of information-theoretic measures, such as Fisher information. In other models, we have to approximate these quantities using Markov Chain Monte Carlo simulations, drastically increasing the computational burden.  Furthermore, by invoking the Hammersley-Clifford theorem \cite{Hammersley}, which states the equivalence between Gibbs random fields (global models) and Markov random fields (local models), we characterize a pairwise isotropic GMRF by the set of local conditional density functions:

\begin{equation}
	p\left( x_{i} | \eta_{i}, \vec{\theta} \right) = \frac{1}{\sqrt{2\pi\sigma^2}}exp\left\{-\frac{1}{2\sigma^{2}} \left[ x_{i} - \mu - \beta \sum_{j \in \eta_{i}} \left( x_{j} - \mu \right) \right]^{2} \right\}
	\label{eq:GMRF}
\end{equation} where $\eta_i$ denotes the the second-order neighborhood system comprised by the 8 nearest neighbors of $x_i$, $\vec{\theta} = (\mu, \sigma^{2}, \beta)$ denotes the vector of model parameters, with $\mu$ and $\sigma^{2}$ being, the expected value (mean) and the variance of the random variables in the lattice, and $\beta$ being the inverse temperature, which encodes the spatial dependence between the variables in the field. Figure \ref{fig:neigh} shows the first, second and third order neighborhood systems defined on a 2D lattice.

Note that if $\beta = 0$, the model degenerates to a regular Gaussian distribution, used to model independent random variables. The main advantage in using the local model is that, by avoiding the joint Gibbs distribution, which has a parametrisation in terms of vectors and matrices, we considerably simplify the parametric space.

\subsection{A curved statistical model}

From mathematical statistics, it is possible to express a likelihood function $p( \mathbf{X}| \vec{\theta} )$ on its natural parameters as:

\begin{equation}
	p(\mathbf{X}| \vec{\theta} ) = exp\left\{\sum_{j=1}^{K}c_{j}( \vec{\theta} )T_{j}\left( \mathbf{X} \right) + d( \vec{\theta} ) + S\left( \mathbf{X} \right) \right\}
\end{equation} where $S\left(\mathbf{X} \right)$ is a function of the observations only, $d(\vec{\theta})$ is a function of the parameters only, $\vec{T} = \left( T_{1}\left(\mathbf{X} \right), T_{2}\left(\mathbf{X} \right), \ldots, T_{k}\left(\mathbf{X} \right) \right)$ denotes the vector of sufficient statistics and $\vec{c} = (c_{1}(\vec{\theta}), c_{2}(\vec{\theta}), \ldots, c_{k}(\vec{\theta}))$ denotes the vector of natural parameters. Let $\mathbf{X} = \{x_{1}, x_{2}, \ldots, x_{n}\}$ be a sample of a pairwise isotropic Gaussian-Markov random field model where $\Delta=8$ represents the number of neighbors. As the number of natural parameters is greater than the number of parameters, isotropic pairwise GMRF's are curved models.
Then, the pseudo-likelihood function can be expressed as:

\begin{align}
	& F\left(\mathbf{X}| \eta_{i}, \vec{\theta} \right) = \left( 2\pi\sigma^2\right)^{-n/2}exp\left\{ -\frac{1}{2\sigma^2}\sum_{i=1}^{n}\left[ \left(x_{i} - \mu \right) - \beta \sum_{j\in\eta_i}\left( x_{j} - \mu \right)  \right]^2  \right\}
\end{align}

\begin{align}
	\nonumber & {} = exp\left\{ -\frac{n}{2}\left[ log(2\pi\sigma^2) + \frac{\mu^2}{\sigma^2} \right] + \frac{\beta\Delta\mu^2 n}{\sigma^2}\left[ 1 - \frac{\beta\Delta}{2} \right] \right\} \times \\ \nonumber 
	& \quad exp\left\{ \left[ \frac{\mu}{\sigma^2}\left(1 - \beta\Delta\right) \right]\sum_{i=1}^{n}x_{i} -\frac{1}{2\sigma^2}\sum_{i=1}^{n}x_{i}^2 + \frac{\beta}{\sigma^2}\sum_{i=1}^{n}\sum_{j\in\eta_i}x_{i}x_{j} \right. \nonumber \\ & \hspace{2cm} \left. - \left[ \frac{\beta\mu}{\sigma^2}(1 - \beta\Delta)\right]\sum_{i=1}^{n}\sum_{j\in\eta_i}x_{j} - \frac{\beta}{2\sigma^2}\sum_{i=1}^{n}\sum_{j\in\eta_i}\sum_{k\in\eta_i}x_{j}x_{k}  \right\} \nonumber
\end{align}

Through direct observation, we can identify the correspondence:

\begin{align}
	\vec{c} & = \left( \left[ \frac{\mu}{\sigma^2}\left(1 - \beta\Delta\right) \right], -\frac{1}{2\sigma^2}, \frac{\beta}{\sigma^2}, -\left[ \frac{\beta\mu}{\sigma^2}(1 - \beta\Delta)\right], - \frac{\beta}{2\sigma^2} \right) \\ \nonumber
	\vec{T} & = \left( \sum_{i=1}^{n}x_{i}, \sum_{i=1}^{n}x_{i}^2, \sum_{i=1}^{n}\sum_{j\in\eta_i}x_{i}x_{j}, \sum_{i=1}^{n}\sum_{j\in\eta_i}x_{j}, \sum_{i=1}^{n}\sum_{j\in\eta_i}\sum_{k\in\eta_i}x_{j}x_{k}  \right) 
\end{align} with $S(\mathbf{X}) = 0$ and 

\begin{equation}
	d(\vec{\theta}) = -\frac{n}{2}\left[ log(2\pi\sigma^2) + \frac{\mu^2}{\sigma^2} \right] + \frac{\beta\Delta\mu^2 n}{\sigma^2}\left[ 1 - \frac{\beta\Delta}{2} \right]
\end{equation}

Furthermore, it is clear that if $\beta = 0$, the pseudo-likelihood function is simplified to the usual likelihood function of the Gaussian model, in which the number of parameters is equal to the number of natural parameters:

\begin{align}
	F\left(\mathbf{X}| \vec{\theta} \right) & = exp\left\{ \frac{\mu}{\sigma^2}\sum_{i=1}^{n}x_{i} -\frac{1}{2\sigma^2}\sum_{i=1}^{n}x_{i}^2 - \frac{n}{2}\left[ log(2\pi\sigma^2) + \frac{\mu^2}{\sigma^2} \right] \right\}
\end{align} where $S(\mathbf{X}) = 0$ and:

\begin{align}
	\vec{c} = \left( \frac{\mu}{\sigma^2}, -\frac{1}{2\sigma^2} \right) \quad
	\vec{T} = \left( \sum_{i=1}^{n}x_{i}, \sum_{i=1}^{n}x_{i}^2 \right) \quad
	d(\vec{\theta}) = -\frac{n}{2}\left[ log(2\pi\sigma^2) + \frac{\mu^2}{\sigma^2} \right]
\end{align}

The previous equations say that the inverse temperature parameter is responsible for making the model curved, as the number of natural parameters becomes greater than the number of parameters. Our goal is to measure how the emergence of the inverse temperature parameter into the model geometrically transforms the underlying parametric space. When the inverse temperature parameter is zero, the parametric space is a surface with constant negative curvature. However, when the inverse temperature increases up to a critical value, the geometry of parametric space suffers significant changes as the system undergoes phase transitions. 

\section{Differential geometry concepts}

\begin{figure}[ht]
	\begin{center}
	\includegraphics[scale=0.4]{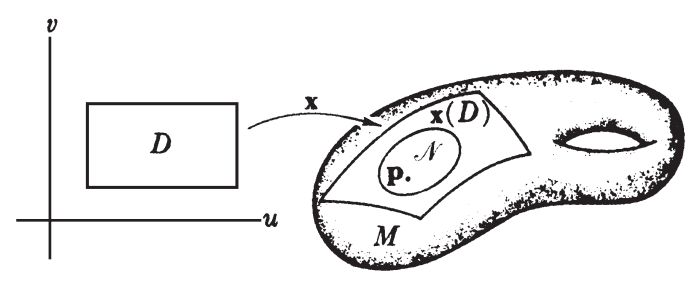}
	\end{center}
	\caption{A surface embedded in $R^3$.}
	\label{fig:surface}
\end{figure}

Differential geometry is a mature research field that is the mathematical background for several areas of physical sciences, remarkably to general relativity, in which it is employed to characterize the geometry of space-time \cite{SeanCarroll}. But, what are the objects of study in differential geometry? Classical differential geometry is the study of local properties of curves and surfaces, where local refers to properties which depend on the behavior of the curve or surface in the neighborhood of a point \cite{Manfredo,Bar}. Before we proceed, it is interesting to give an intuition behind the concept of surface: a surface in $R^3$ can be constructed by taking pieces of a plane, deforming them and arranging them in a way that the resulting shape has no sharp points, edges or self-intersections so that we can assign a tangent plane to every point of the surface \cite{Manfredo,Bar,Oneill,Pressley}. In other words, a surface in $R^3$ looks like an open subset of $R^2$ which has been smoothly deformed. An example of surface is the parametric space of a Gaussian random variable with a probability density function $p(x; \vec{\theta})$, in which $\vec{\theta} = (\mu, \sigma^2)$, where $\mu$ is the mean and $\sigma^2$ is the variance. Figure \ref{fig:surface}, originally from O'Neill's book on differential geometry \cite{Oneill}, illustrates an arbitrary surface embedded in the $R^3$ ambient space.

\begin{definition}[Surface]
	A surface in $R^3$ is a subset $M \subset R^3$ such that for each point $p \in M$ there exists a proper patch in $M$ whose image contains a neighborhood of $p$ in $M$, where a proper patch is a one-to-one mapping of an open set $D \subset R^2$ into $R^3$.
\end{definition}

In order to adapt the calculus of the plane $R^2$ to an arbitrary surface, a fundamental concept is the tangent plane. The intuition is that as lines are the simplest curves, the simplest surfaces are planes, so it is possible to approximate very complicated surfaces by simple planes. 

\begin{definition}[Tangent plane]
	Let $M \subset R^3$ be a regular surface and $p \in M$ an arbitrary point. A vector $\vec{v}$ is tangent to $M$ at $p$ provided there exists an $\epsilon > 0$ and a smooth parametrized curve $\alpha: (-\epsilon, \epsilon) \rightarrow M$ with $\alpha(0) = p$ and $\alpha'(0) = \vec{v}$. The set of all tangent vectors at $p$ defines the tangent plane in $p$, denoted by $T_pM$. 
\end{definition}

\begin{figure}[ht]
	\begin{center}
	\includegraphics[scale=0.3]{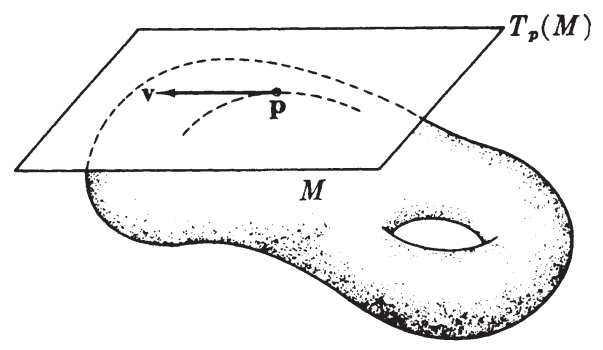}
	\end{center}
	\caption{A surface $M$ and the tangent plane $T_pM$.}
	\label{fig:tangent}
\end{figure}

Figure \ref{fig:tangent}, originally from O'Neill's book \cite{Oneill}, illustrates an arbitrary surface and the tangent plane at an arbitrary point $p \in M$.

The first thing a surface inhabitant needs to do in geometric terms is to measure the distance between two points and the angle between two vectors belonging to a tangent space. This is different from the distance between these points as measured by someone that lives in the the ambient space, since, often, the straight line between the points in $R^3$ will not be contained in the surface. The mathematical object that allows one to compute lengths on a surface, and also angles and areas, is the first fundamental form of the surface \cite{Pressley,Woodward}.

\begin{definition}[First fundamental form]
	Let $p$ be an arbitrary point of a surface $M$. The first fundamental form of $M$ at $p$ associate to tangent vectors $\vec{v}, \vec{w} \in T_pM$ the scalar:
	\begin{equation}
		\langle \vec{w}, \vec{z} \rangle_{p, M} = \vec{w} \cdot \vec{z}
	\end{equation}
	In other words, the first fundamental form allows us to compute dot products in the tangent plane.
\end{definition}

Suppose $x(u, v)$ defines a local parametrisation of a surface $M$. Then, any tangent vector to $M$ at a point $p$ can be expressed uniquely as a linear combination of $\vec{x}_u$ and $\vec{x}_v$, obtained by differentiating $x(u, v)$ with respect to $u$ and $v$, respectively. Hence, we can write: 

\begin{align}
	\vec{w} = \lambda_1 \vec{x}_u + \lambda_2 \vec{x}_v \\
	\vec{z} = \mu_1 \vec{x}_u + \mu_2 \vec{x}_v 
\end{align} and the inner product becomes:

\begin{align}
	\langle \vec{w}, \vec{z} \rangle_{p, M} & = (\lambda_1 \vec{x}_u + \lambda_2 \vec{x}_v) \cdot (\mu_1 \vec{x}_u + \mu_2 \vec{x}_v) \\ \nonumber & = \lambda_1\mu_1 E + (\lambda_1\mu_2 + \lambda_2\mu_1) F + \lambda_2\mu_2 G 
\end{align} where

\begin{align}
	E = \vec{x}_u \cdot \vec{x}_u \qquad F = \vec{x}_u \cdot \vec{x}_v \qquad G = \vec{x}_v \cdot \vec{x}_v 
\end{align}

The functions $E$, $F$ and $G$ are called the coefficients of the first fundamental form of the surface. This structure is important because it enables the computation of arc lengths of curves on the surface and the areas of regions on the surface. Note that an infinitesimal displacement in the surface $ds$ may be expressed in terms of the coefficients of the first fundamental form as:

\begin{equation}
	ds^2 = E du^2 + 2 F du dv + G dv^2
\end{equation}

In many applications, the first fundamental form is often written in the modern notation of the metric tensor:

\begin{equation}
	\langle \vec{w}, \vec{z} \rangle_{p, M} = \vec{w}^T  \begin{bmatrix} 
	E & F \\
	F & G \\
	\end{bmatrix}  \vec{z}
\end{equation} where $F = 0$ implies in orthogonal coordinate systems. Note that if we have $E = G = 1$ and $F = 0$, we achieve the Euclidean geometry. It is usual to denote the first fundamental form by $I$.

The first fundamental form relates to the study of intrinsic properties of surfaces. However, there are several extrinsic properties that are relevant for the complete characterization of a surface, such as the Gaussian curvature and the mean curvature. In summary, to measure how a surface is curving at a point $p$, we need to measure the rate of change of change of the unit normal vector at $p$. This discussion involves the definition of the second fundamental form of a surface \cite{Pressley}.

\begin{figure}[ht]
	\begin{center}
	\includegraphics[scale=0.15]{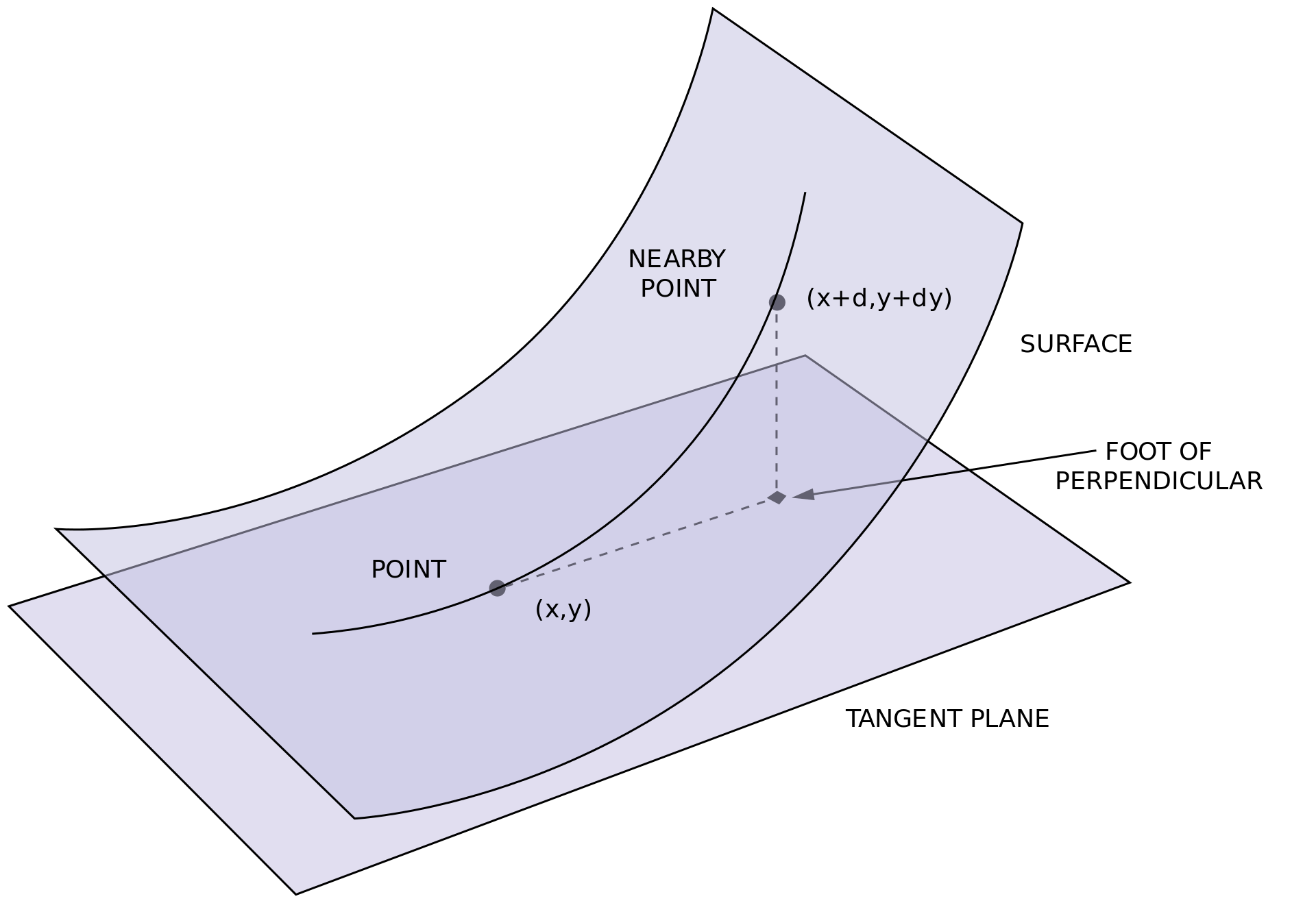}
	\end{center}
	\caption{The notion of curvature is related to how fast the surface escapes the tangent plane.}
	\label{fig:curvature}
\end{figure}

\begin{definition}[Second fundamental form]
	Let $x(u, v)$ be a local parametrisation of a surface $M$ with standard unit normal $\vec{N}$. As the local coordinates $(u, v)$ change to $(u+\Delta u, v+\Delta v)$ the surface moves away from the tangent plane by the distance $d$ defined by:
\begin{equation}
	d = ( x(u+\Delta u, v+\Delta v) - x(u,v) ) \cdot \vec{N}
\end{equation} as illustrates Figure \ref{fig:curvature}.
\end{definition}

By a Taylor expansion, it is possible to express $x(u+\Delta u, v+\Delta v) - x(u,v)$ as:

\begin{equation}
	x_u \Delta u + x_v \Delta v + \frac{1}{2}\left( x_{uu}(\Delta u)^2 + 2 x_{uv}\Delta u \Delta v + x_{vv}(\Delta v)^2 \right) + r(\Delta u, \Delta v)
\end{equation} where the remainder function $r$ becomes negligible when $(\Delta u)^2 + (\Delta v)^2$ tends to zero (infinitesimal displacement). Note that as both $x_u$ and $x_v$ belong to the tangent place, they are orthogonal to the normal vector $\vec{n}$, which means that the deviation $d$ can be expressed as:

\begin{equation}
	d = \frac{1}{2}\left( L (\Delta u)^2 + 2M \Delta u \Delta v + N (\Delta v)^2 \right)
\end{equation} where

\begin{align}
	L = x_{uu} \cdot \vec{n} \qquad M = x_{uv} \cdot \vec{n} \qquad N = x_{vv} \cdot \vec{n}
\end{align}

When $\Delta u \rightarrow 0$ and $\Delta v \rightarrow 0$, the resulting expression:

\begin{equation}
	L du^2 + 2M du dv + N dv^2
\end{equation} defines the second fundamental form of the surface, which, often, is denoted by $II$ and organized as a square matrix:

\begin{equation}
	II = \begin{bmatrix} 
	L & M \\
	M & N \\
	\end{bmatrix}
\end{equation}

In order to compute the surface curvatures, it is necessary to obtain the expressions of the second fundamental form and of the differential of the Gauss map in a coordinate system. In computational terms, this can be done through the definition of the shape operator \cite{Manfredo}.

\begin{definition}[Shape operator]
	Let $M$ be a surface with first fundamental form $I$ and second fundamental form $II$. Then, the shape operator $P$ can be computed by:
\begin{equation}
	P = -(II)(I)^{-1} = - \begin{bmatrix} 
	L & M \\
	M & N \\
	\end{bmatrix} \begin{bmatrix} 
	E & F \\
	F & G \\
	\end{bmatrix}^{-1}
\end{equation}	
\end{definition}

The shape operator encodes relevant information about the curvature of surfaces, being a powerful mathematical tool for geometric analysis. From the shape operator, we can obtain the Gaussian, mean and principal curvatures.

\begin{definition}
	Let $M$ be a surface and $P$ its shape operator. It can be shown that \cite{Manfredo}:
	\begin{enumerate}
		\item The Gaussian curvature, $K$, is the determinant of the shape operator $P$.
		\item The mean curvature, $H$, is the trace of the shape operator $P$.
		\item The principal curvatures are the eigenvalues of the shape operator $P$.
	\end{enumerate}
\end{definition}

In the next sections, we will discuss how to derive the first and second fundamental forms of Gaussian random field manifolds by means of the first-order and second-order Fisher information matrices. Undoubtedly, one of the most important findings of Carl Friedrich Gauss is that the Gaussian curvature is unchanged when the surface is bent without stretching, which is formally stated by the remarkable \emph{Theorema Egregium} \cite{Pressley}. In other words, this result shows that the curvatures of a surface can be computed using only the first fundamental form. The implication of this theorem to our study is related to the amount of stretching the parametric space of Gaussian random fields suffers when it bends depends whether the inverse temperature is increasing or decreasing.


\section{Fisher information}

Information geometry is a recent research field that combines differential geometry and information theory to study intrinsic properties of the parametric spaces of several statistical models \cite{Amari,Nielsen}. According to this theory, the first-order Fisher information matrix is the metric tensor that equips the underlying parametric space of a statistical manifold (first fundamental form), whereas the second-order Fisher information matrix is the second fundamental form of the parametric space. It allows the application of the same mathematical tools used in the study deterministic physical phenomena, such as general relativity, into the macroscopic description of random systems \cite{Quantum}.

In practice, the metric tensor allows us to express the square of an infinitesimal displacement in the manifold, $ds^2$, as a function of an infinitesimal displacement in the tangent space. Assuming a surface $M$ and a matrix notation, we have:

\begin{equation}
	ds^2 = \begin{bmatrix} du & dv \end{bmatrix} \begin{bmatrix} E & F \\ F & G \end{bmatrix}\begin{bmatrix} du \\ dv \end{bmatrix} = E du^2 + 2F du dv + G dv^2
\end{equation} where the matrix of coefficients $E$, $F$, e $G$ is the first fundamental form. If this matrix is positive definite, we have a Riemannian manifold. Note that in the Euclidean case, the metric tensor is the identity matrix (since the the space is flat), and we have the known Pythagorean relation $ds^2 = du^2 + dv^2$. 

The second-order Fisher information matrix (second fundamental form) encodes information about the curvature of the manifold and it is used in the definition of the shape operator, whose determinant, trace and eigenvalues correspond to the Gaussian, mean and principal curvatures. Our goal is to measure the variations in the geometrical structure of the parametric space of pairwise isotropic Gaussian-Markov random fields along the evolution of these systems under phase transitions.

\begin{definition}[First-order Fisher information matrix]
	Let $p(X;\vec{\theta})$ be a probability density function where $\vec{\theta} = (\theta_1, \ldots, \theta_n) \in \Theta$ is the vector of parameters. The first-order Fisher information matrix, which is the first fundamental form of underlying parametric space, is defined as:
\begin{equation}
	\left\{ I(\vec{\theta}) \right\}_{ij} = E\left[ \left(\frac{\partial}{\partial\theta_i} log~p(X; \vec{\theta}) \right)\left(\frac{\partial}{\partial\theta_j} log~p(X;\vec{\theta}) \right) \right], \text{~~~~ for } i,j=1,\ldots,n
\end{equation}
\end{definition}

It has been shown that, under certain regularity conditions, the information equality is valid, which means that both first and second-order Fisher information matrices are identical \cite{Bickel,Lehman,Casella}.

\begin{definition}[Second-order Fisher information matrix]
	Let $p(X;\vec{\theta})$ be a probability density function where $\vec{\theta} = (\theta_1, \ldots, \theta_n) \in \Theta$ is the vector of parameters. The second-order Fisher information matrix, which is the second fundamental form of underlying parametric space, is defined as:
\begin{equation}
	\left\{ II(\vec{\theta}) \right\}_{ij} = -E\left[ \frac{\partial^2}{\partial\theta_i \partial\theta_j} log~p(X; \vec{\theta}) \right], \text{~~~~ for } i,j=1,\ldots,n
\end{equation}
\end{definition}

\subsection{Considerations about the information equality}

In the following, we provide a brief discussion about the information equality condition, which is a valid property for several probability density function belonging to the exponential family when the observations are independent \cite{Silvey}. Let $X$ be a random variable with a probability density function $p(X;\vec{\theta})$. First, note that:

\begin{align}
	E\left[ \frac{\partial^2}{\partial\theta_i \partial\theta_j}log~p(x; \vec{\theta}) \right]  & = E\left[ \frac{\partial}{\partial\theta_j}\left(\frac{\partial}{\partial\theta_i} log~p(x; \vec{\theta}) \right)  \right] \nonumber \\ & = E\left[ \frac{\partial}{\partial\theta_j}\left(\frac{1}{p(x; \vec{\theta})} \frac{\partial}{\partial\theta_i} p(x; \vec{\theta}) \right)  \right]
\end{align}

By direct application of the product rule, we have:

\begin{align}
	& \nonumber E\left[ \frac{\partial}{\partial\theta_j}\left(\frac{1}{p(x; \vec{\theta})} \frac{\partial}{\partial\theta_i} p(x; \vec{\theta}) \right)  \right] = E \left[ -\frac{1}{p(x; \vec{\theta})^2} \frac{\partial}{\partial\theta_i} p(x; \vec{\theta}) \frac{\partial}{\partial\theta_j} p(x; \vec{\theta}) \right. \\ \nonumber & \qquad\qquad\qquad\qquad\qquad\qquad\qquad\qquad\qquad\qquad\qquad \left. + \frac{1}{p(x; \vec{\theta})} \frac{\partial^2}{\partial\theta_i \partial\theta_j} p(x; \vec{\theta}) \right] \\ \nonumber & = -E\left[ \left( \frac{1}{p(x; \vec{\theta})} \frac{\partial}{\partial\theta_i}p(x; \vec{\theta})  \right) \left( \frac{1}{p(x; \vec{\theta})} \frac{\partial}{\partial\theta_i}p(x; \vec{\theta})  \right) \right] \\ & \qquad\qquad\qquad\qquad\qquad\qquad\qquad\qquad\qquad + E\left[ \frac{1}{p(x; \vec{\theta})} \frac{\partial^2}{\partial\theta_i \partial\theta_j} p(x; \vec{\theta}) \right]
	\label{eq:34}
\end{align}

Using the definition of the expectation operator, the second term of equation \eqref{eq:34} can be expressed as:

\begin{align}
	E\left[ \frac{1}{p(x; \vec{\theta})} \frac{\partial^2}{\partial\theta_i \partial\theta_j} p(x; \vec{\theta}) \right] & = \int p(x; \vec{\theta}) \frac{1}{p(x; \vec{\theta})} \frac{\partial^2}{\partial\theta_i \partial\theta_j} p(x; \vec{\theta}) dx \nonumber \\ & = \int \frac{\partial^2}{\partial\theta_i \partial\theta_j} p(x; \vec{\theta}) dx
\end{align}

Under certain regularity conditions it is possible to interchange the integration and differentiation operators:

\begin{align}
	\int \frac{\partial^2}{\partial\theta_i \partial\theta_j} p(x; \vec{\theta}) dx = \frac{\partial^2}{\partial\theta_i \partial\theta_j} \int p(x; \vec{\theta}) dx = \frac{\partial^2}{\partial\theta_i \partial\theta_j} 1 = 0
\end{align}

Note also that the arguments in the expectation in the first term of equation (\ref{eq:34}) can be rewritten as:

\begin{equation}
	\frac{1}{p(x; \vec{\theta})} \frac{\partial}{\partial\theta_i} p(x; \vec{\theta}) = \frac{\partial}{\partial\theta_i} log~p(x; \vec{\theta})
\end{equation} which finally leads to the equality:

\begin{equation}
	E\left[ \left( \frac{\partial}{\partial\theta_i} log~p(x; \vec{\theta}) \right) \left( \frac{\partial} {\partial\theta_j} log~p(x; \vec{\theta}) \right) \right] = -E\left[ \frac{\partial^2}{\partial\theta_i \partial\theta_j} log~p(x;\vec{\theta}) \right]
\end{equation}

In geometric terms, it means that the elements of the first fundamental form ($I$) are equal to the elements of the second fundamental form ($II$), which means constant Gaussian curvature. We will see that this is not the case in pairwise isotropic Gaussian-Markov random fields.

\subsection{Derivation of the first fundamental form}

In this Section, we provide the mathematical derivation of the components of the first fundamental form. As the parametric space is a 3D manifold, the first-order Fisher information matrix has the shape:

\begin{equation}
	I(\vec{\theta}) = \left( \begin{array}{ccc}
	A & B & C \\ 
	D & E & F \\ 
	G & H & I
	\end{array} \right)
\end{equation}

We will begin with the first component of the matrix, which involves the derivatives with respect to the $\mu$ parameter. Note that the first component of $I(\vec{\theta})$ is:

\begin{equation}
	A = I_{\mu\mu}(\vec{\theta}) = E\left[ \left(\frac{\partial}{\partial\mu} log~p(X; \vec{\theta}) \right)\left(\frac{\partial}{\partial\mu} log~p(X; \vec{\theta}) \right) \right]
\end{equation} where $p(X; \vec{\theta})$ is the replaced by the local conditional density function of the Gaussian random field, given by equation \eqref{eq:GMRF}. The computation of the derivatives leads to:

\begin{align}
	A & = E\left\{ \frac{1}{\sigma^2}\left(1 - \beta\Delta \right)^2 \frac{1}{\sigma^2}\left[ \left(x_i - \mu \right) - \beta \sum_{j\in\eta_i}\left(x_j - \mu \right) \right]^2 \right\} \label{eq:mu_mu_1} 
\end{align}

Expanding the square, we have:

\begin{align}
	A & = \frac{1}{\sigma^2}\left(1 - \beta\Delta \right)^2 E\left\{ \frac{1}{\sigma^2} \left[ \left(x_i - \mu\right)^2 - 2\beta\sum_{j\in\eta_i}\left( x_i - \mu \right)\left( x_j - \mu \right) \right. \right. \\ \nonumber & \hspace{4cm} \left. \left. + \beta^2 \sum_{j\in\eta_i}\sum_{k\in\eta_i}\left( x_j - \mu \right)\left(x_k - \mu  \right) \right] \right\} 
\end{align} 

And simplifying the expected values, we reach:

\begin{equation}
	A = \frac{\left(1 - \beta\Delta \right)^2}{\sigma^2} \left[ 1 - \frac{1}{\sigma^2}\left(  2\beta\sum_{j\in\eta_i}\sigma_{ij} - \beta^2\sum_{j\in\eta_i}\sum_{k\in\eta_i}\sigma_{jk} \right) \right]
\end{equation} where $\Delta$ is the cardinality of the neighborhood system ($\Delta = 8$ in a second-order system), $\sigma_{ij}$ is the covariance between the central variable $x_i$ and one of its neighbors $x_j \in \eta_i$ and $\sigma_{jk}$ is the covariance between two variables $x_j$ and $x_k$ belonging to the neighborhood $\eta_i$. The second component of the first fundamental form is:

\begin{equation}
	B = I_{\mu\sigma^2}(\vec{\theta}) = E\left[ \left(\frac{\partial}{\partial\mu} log~p(X; \vec{\theta}) \right)\left(\frac{\partial}{\partial\sigma^2} log~p(X; \vec{\theta}) \right) \right]
\end{equation} which leads to:

\begin{align}
	\label{eq:g12}
	B & = \frac{(1 - \beta\Delta)}{2\sigma^6}E\left\{\left[ \left( x_i - \mu \right) - \beta\sum_{j\in\eta_i}\left(x_j - \mu \right) \right]^3 \right\} \\ \nonumber & \hspace{3cm} -\frac{(1 - \beta\Delta)}{2\sigma^4}E\left\{ \left( x_i - \mu \right) - \beta\sum_{j\in\eta_i}\left(x_j - \mu \right) \right\}	
\end{align}

Note that second term of equation \eqref{eq:g12} is zero, since:

\begin{equation}
	E\left[ x_i - \mu \right] - \beta\sum_{j\in\eta_i}E\left[ x_j - \mu\right] = 0 - 0 = 0
\end{equation} and the expansion of the first term of equation \eqref{eq:g12} leads to:

\begin{align}
	\label{eq:g12_exp}
	& E\left\{\left[ \left( x_i - \mu \right) - \beta\sum_{j\in\eta_i}\left(x_j - \mu \right) \right]^3 \right\} = E\left[ \left( x_i - \mu \right)^3 \right] \\ \nonumber & - 3\beta\sum_{j\in\eta_i}E\left[ (x_i - \mu) (x_i - \mu) (x_j - \mu) \right] \\ \nonumber & + 3\beta^2 \sum_{j\in\eta_i}\sum_{k\in\eta_i}E\left[ (x_i - \mu) (x_j - \mu) (x_k - \mu) \right] \\ \nonumber & - \beta^3 \sum_{j\in\eta_i}\sum_{k\in\eta_i}\sum_{l\in\eta_i}E\left( (x_j - \mu) (x_k - \mu) (x_l - \mu) \right]
\end{align}
	
The first term of \eqref{eq:g12_exp} is zero for Gaussian random variables, since every central moment of odd order is null. According to the Isserlis' theorem \cite{isserlis1918}, it is trivial to see that in fact all the third order cross terms are null, therefore, $B=0$. The third component of the first fundamental form is:

\begin{equation}
	C = I_{\mu\beta}(\vec{\theta}) = E\left[ \left(\frac{\partial}{\partial\mu} log~p(X; \vec{\theta}) \right)\left(\frac{\partial}{\partial\beta} log~p(X; \vec{\theta}) \right) \right]
\end{equation}

Plugging the local conditional density function and doing some basic algebra, we reach:

\begin{align}
	C = I_{\mu\beta}(\vec{\theta}) & = \frac{(1 - \beta\Delta)}{\sigma^4}\Bigg\{ E\left[ (x_i - \mu) (x_i - \mu) (x_j - \mu) \right]  \\ \nonumber & - 2\beta\sum_{j\in\eta_i}\sum_{k\in\eta_i}E\left[ (x_i - \mu) (x_j - \mu) (x_k - \mu) \right] \\ \nonumber & + \beta^2 \sum_{j\in\eta_i}\sum_{k\in\eta_i}\sum_{l\in\eta_i}E\left[ (x_j - \mu) (x_k - \mu) (x_l - \mu) \right] \Bigg\}
\end{align}

Once again, all the  third order moments are zero by the Isserlis's theorem, resulting in $C=0$. For the next component, by the symmetry of the metric tensor, $D=B=0$. In order to calculate fifth component of the first fundamental form, we have to compute:

\begin{equation}
	E = I_{\sigma^2 \sigma^2}(\vec{\theta}) = E\left[ \left(\frac{\partial}{\partial\sigma^2} log~p(X; \vec{\theta}) \right)\left(\frac{\partial}{\partial\sigma^2} log~p(X; \vec{\theta}) \right) \right]
\end{equation} which is given by:

\begin{align}
	E & = E\left\{ \left[ -\frac{1}{2\sigma^2} + \frac{1}{2\sigma^4}\left( x_i - \mu - \beta\sum_{j\in\eta_i}(x_j - \mu) \right) \right]^2 \right\} \\ \nonumber & = \frac{1}{4\sigma^4} - \frac{1}{2\sigma^6}E\left\{ \left[ (x_i - \mu) - \beta\sum_{j\in\eta_i}(x_j - \mu) \right]^2 \right\} \\ \nonumber & \hspace{1cm} + \frac{1}{4\sigma^8}E\left\{ \left[ (x_i - \mu) - \beta\sum_{j\in\eta_i}(x_j - \mu) \right]^4 \right\}
\end{align}

Note that the first expectation leads to the following equality:

\begin{equation}
	E\left\{ \left[ (x_i - \mu) - \beta\sum_{j\in\eta_i}(x_j - \mu) \right]^2 \right\} =  \sigma^2 - 2\beta\sum_{j\in\eta_i}\sigma_{ij} + \beta^2 \sum_{j\in\eta_i}\sum_{k\in\eta_i}\sigma_{jk} 
\end{equation}

For the second expectation, we have:

\begin{align}
	& E\left\{ \left[ (x_i - \mu) - \beta\sum_{j\in\eta_i}(x_j - \mu) \right]^4 \right\} = E\left[ (x_i - \mu)^4 \right] \\ \nonumber & \hspace{1cm} - 4\beta\sum_{j\in\eta_i}E\left[ (x_i - \mu)^3 (x_j - \mu) \right] \nonumber \\ & \hspace{1cm} + 6\beta^2 \sum_{j\in\eta_i}\sum_{k\in\eta_i}E\left[ (x_i - \mu)^2 (x_j - \mu) (x_k - \mu) \right] \\ \nonumber & \hspace{1cm} - 4\beta^3 \sum_{j\in\eta_i}\sum_{k\in\eta_i}\sum_{l\in\eta_i} E\left[(x_i - \mu) (x_j - \mu) (x_k - \mu) (x_l - \mu) \right] \\ \nonumber & \hspace{1cm} + \beta^4 \sum_{j\in\eta_i}\sum_{k\in\eta_i}\sum_{l\in\eta_i}\sum_{m\in\eta_i}E\left[(x_j - \mu) (x_k - \mu) (x_l - \mu) (x_m - \mu) \right]
\end{align} leading to five different expectation terms. We invoke the Isserlis' theorem for Gaussian random variables to express higher order moments in terms of second-order moments. Hence, after some algebraic manipulations, we have:

\begin{align}
	\label{eq:sigma_sigma_1}
	E & = \frac{1}{2\sigma^4} - \frac{1}{\sigma^6}\left[ 2\beta\sum_{j\in\eta_i}\sigma_{ij} - \beta^2 \sum_{j\in\eta_i}\sum_{k\in\eta_i}\sigma_{jk} \right] \\ \nonumber & + \frac{1}{\sigma^8}\left[ 3\beta^2 \sum_{j\in\eta_i}\sum_{k\in\eta_i}\sigma_{ij}\sigma_{ik} - \beta^3 \sum_{j\in\eta_i}\sum_{k\in\eta_i}\sum_{l\in\eta_i}\left( \sigma_{ij}\sigma_{kl} + \sigma_{ik}\sigma_{jl} + \sigma_{il}\sigma_{jk} \right) \right. \\ \nonumber & \hspace{1cm} \left. + \beta^4 \sum_{j\in\eta_i}\sum_{k\in\eta_i}\sum_{l\in\eta_i}\sum_{m\in\eta_i}\left( \sigma_{jk} \sigma_{lm} + \sigma_{jl}\sigma_{km} + \sigma_{jm}\sigma_{kl} \right)  \right] 
\end{align}

The sixth component of the first fundamental form is given by:

\begin{equation}
	F = I_{\sigma^2 \beta}(\vec{\theta}) = E\left[ \left(\frac{\partial}{\partial\sigma^2} log~p(X; \vec{\theta}) \right)\left(\frac{\partial}{\partial\beta} log~p(X; \vec{\theta}) \right) \right]
\end{equation} which can be computed as:

\begin{align}
	 F = & E\left\{ \left[ -\frac{1}{2\sigma^2} + \frac{1}{2\sigma^4}\left( (x_i - \mu) - \beta\sum_{j\in\eta_i}(x_j - \mu)  \right)^2 \right] \times \right. \\ \nonumber & \hspace{2cm} \left. \left[ \frac{1}{\sigma^2}\left((x_i - \mu) - \beta\sum_{j\in\eta_i}(x_j - \mu)  \right)\left( \sum_{j\in\eta_i}(x_j - \mu) \right) \right] \right\}
\end{align}

\begin{align}	 
	\nonumber {} & = -\frac{1}{2\sigma^4} E\left\{  \left[ (x_i - \mu) - \beta\sum_{j\in\eta_i}(x_j - \mu) \right]\left[ \sum_{j\in\eta_i}(x_j - \mu) \right] \right\} \\ \nonumber & \hspace{2cm} + \frac{1}{2\sigma^6}E\left\{\left[ (x_i - \mu) - \beta\sum_{j\in\eta_i}(x_j - \mu) \right]^3 \left[ \sum_{j\in\eta_i}(x_j - \mu) \right]  \right\}
\end{align}

By computing the first expectation, we have:

\begin{equation}
	E\left\{  \left[ (x_i - \mu) - \beta\sum_{j\in\eta_i}(x_j - \mu) \right]\left[ \sum_{j\in\eta_i}(x_j - \mu) \right] \right\} = \sum_{j\in\eta_i}\sigma_{ij} - \beta\sum_{j\in\eta_i}\sum_{k\in\eta_i}\sigma_{jk}
\end{equation}

The expansion of the second expectation leads to:

\begin{align}
	& E\left\{\left[ (x_i - \mu) - \beta\sum_{j\in\eta_i}(x_j - \mu) \right]^3 \left[ \sum_{j\in\eta_i}(x_j - \mu) \right]  \right\} = \\ \nonumber & E\left\{ \left[ \sum_{j\in\eta_i}(x_j - \mu) \right] \left[ (x_i - \mu)^3 - 3\beta\sum_{j\in\eta_i}(x_i - \mu)^2 (x_j - \mu) \right. \right. \\ \nonumber \\ \nonumber & \hspace{4cm} \left. \left. + 3\beta^2 \sum_{j\in\eta_i}\sum_{k\in\eta_i}(x_i - \mu)(x_j - \mu)(x_k - \mu) \right. \right. \\ \nonumber  & \hspace{5cm} \left. \left. -\beta^3 \sum_{j\in\eta_i}\sum_{k\in\eta_i}\sum_{l\in\eta_i}(x_j - \mu)(x_k - \mu)(x_l - \mu) \right] \right\}
\end{align}

Again, by direct application of the Isserlis' theorem to express higher-order cross moments in terms of second-order moments and after some simplifications, we have:

\begin{align}
	\label{eq:sigma_beta_1}
	F & = \frac{1}{\sigma^4}\left[ \sum_{j\in\eta_i}\sigma_{ij} - \beta\sum_{j\in\eta_i}\sum_{k\in\eta_i}\sigma_{jk} \right] \\ \nonumber & - \frac{1}{2\sigma^6}\left[ 6\beta\sum_{j\in\eta_i}\sum_{k\in\eta_i}\sigma_{ij}\sigma_{ik} - 3 \beta^2 \sum_{j\in\eta_i}\sum_{k\in\eta_i}\sum_{l\in\eta_i}\left( \sigma_{ij}\sigma_{kl} + \sigma_{ik}\sigma_{jl} + \sigma_{il}\sigma_{jk} \right) \right. \\ \nonumber & \hspace{3cm} \left. + \beta^3 \sum_{j\in\eta_i}\sum_{k\in\eta_i}\sum_{l\in\eta_i}\sum_{m\in\eta_i} \left( \sigma_{jk}\sigma_{lm} + \sigma_{jl}\sigma_{km} + \sigma_{jm}\sigma_{kl} \right) \right]
\end{align}

It is straightforward to see that $G=C=0$ and $G=F$, since the first fundamental form is symmetric. Finally, the last component is defined as:

\begin{equation}
	I = I_{\beta\beta}(\vec{\theta}) = E\left[ \left(\frac{\partial}{\partial\beta} log~p(X; \vec{\theta}) \right)\left(\frac{\partial}{\partial\beta} log~p(X; \vec{\theta}) \right) \right]
\end{equation} which is given by:

\begin{align}
	I & = \frac{1}{\sigma^4}E\left\{ \left[ (x_i - \mu) - \beta\sum_{j\in\eta_i}(x_j - \mu) \right]^2 \left[ \sum_{j\in\eta_i}(x_j - \mu) \right]^2  \right\} \\ \nonumber & = \frac{1}{\sigma^4} E\left\{ \left[ (x_i - \mu)^2 - 2\beta \sum_{j\in\eta_i} (x_i - \mu)(x_j - \mu) + \beta^2 \sum_{j\in\eta_i}\sum_{k\in\eta_i} (x_j - \mu)(x_k - \mu) \right] \times \right. \\ \nonumber & \left. \hspace{5cm} \left[ \sum_{j\in\eta_i}\sum_{k\in\eta_i} (x_j - \mu)(x_k - \mu) \right] \right\} \\ \nonumber & = \frac{1}{\sigma^4} E \left\{ \sum_{j\in\eta_i}\sum_{k\in\eta_i}(x_i - \mu)(x_i - \mu)(x_j - \mu)(x_k - \mu) \right. \\ \nonumber & \hspace{2cm} \left. - 2\beta\sum_{j\in\eta_i}\sum_{k\in\eta_i} \sum_{l\in\eta_i}(x_i - \mu)(x_j - \mu)(x_k - \mu)(x_l - \mu) \right. \\ \nonumber & \hspace{3cm} \left. + \beta^2 \sum_{j\in\eta_i} \sum_{k\in\eta_i} \sum_{l\in\eta_i} \sum_{m\in\eta_i}(x_j - \mu)(x_k - \mu) (x_l - \mu) (x_m - \mu)  \right\}
\end{align}
 
Once again, by using the Isserlis' formula and some algebra, we have:

\begin{align}
	\label{eq_beta_beta_1}
	I = \frac{1}{\sigma^2}\sum_{j\in\eta_i} \sum_{k\in\eta_i} \sigma_{jk} & + \frac{1}{\sigma^4} \left[ 2 \sum_{j\in\eta_i} \sum_{k\in\eta_i} \sigma_{ij} \sigma_{ik} \right. \\ \nonumber & \left. - 2\beta \sum_{j\in\eta_i} \sum_{k\in\eta_i} \sum_{l\in\eta_i} \left( \sigma_{ij}\sigma_{kl} + \sigma_{ik}\sigma_{jl} + \sigma_{il}\sigma_{jk} \right) \right. \\ \nonumber & \left. + \beta^2 \sum_{j\in\eta_i} \sum_{k\in\eta_i} \sum_{l\in\eta_i} \sum_{m\in\eta_i} \left( \sigma_{jk}\sigma_{lm} + \sigma_{jl}\sigma_{km} + \sigma_{jm}\sigma_{kl} \right) \right]
\end{align} concluding that the first fundamental form has the following structure:

\begin{equation}
	I(\vec{\theta}) = \left( \begin{array}{ccc}
	A & 0 & 0 \\ 
	0 & E & F \\ 
	0 & F & I
	\end{array} \right)
\end{equation} where the non-zero elements are used to define how we compute an infinitesimal displacement in the manifold (parametric space) around the point $\vec{p} = (\mu, \sigma^2, \beta)$:

\begin{align}
	(ds)^2 & = \begin{bmatrix} d\mu & d\sigma^2 & d\beta \end{bmatrix}  \begin{bmatrix} A & 0 & 0 \\ 
	0 & E & F \\ 
	0 & F & I \end{bmatrix} \begin{bmatrix} d\mu \\ d\sigma^2 \\ d\beta \end{bmatrix} \\ \nonumber & = A (d\mu)^2 + E (d\sigma^2)^2 + I (d\beta)^2 + 2 F (d\beta)(d\sigma^2)
\end{align}

\subsection{Derivation of the second fundamental form}

In the following, we proceed with the derivation of the second fundamental form of the model, by computing the elements of the second-order Fisher information matrix. As the parametric space is a 3D manifold, the second fundamental form has the shape:

\begin{equation}
	II(\vec{\theta}) = \left( \begin{array}{ccc}
	L & M & N \\ 
	O & P & Q \\ 
	R & S & T
	\end{array} \right)
\end{equation}

The first component of the second fundamental form $II(\vec{\theta})$ is:

\begin{equation}
	L = II_{\mu\mu} (\vec{\theta}) = -E\left[ \frac{\partial^2}{\partial\mu^2} log~p(X; \vec{\theta}) \right]
\end{equation} which is given by:

\begin{align}
	L = -\frac{\left( 1 - \beta\Delta \right)}{\sigma^2}E\left\{ \frac{\partial}{\partial\mu}\left[ (x_i - \mu) - \beta\sum_{j\in\eta_i}(x_j - \mu) \right] \right\}	= \frac{1}{\sigma^2}\left( 1 - \beta\Delta \right)^2
\end{align} where $\Delta=8$ is the size of the neighborhood system. The second component is defined by:

\begin{equation}
	M = II_{\mu\sigma^2} (\vec{\theta}) = -E\left[ \frac{\partial^2}{\partial\mu \partial\sigma^2} log~p(X; \vec{\theta}) \right]
\end{equation} resulting in

\begin{align}
	M = \frac{1}{\sigma^4}(1 - \beta\Delta) E\left[ (x_i - \mu) - \beta\sum_{j\in\eta_i}(x_j - \mu) \right] = \frac{1}{\sigma^4}(1 - \beta\Delta)\left[ 0 - 0 \right] = 0
\end{align}

The third component of the second fundamental form is also zero:

\begin{align}
	N = II_{\mu\beta} (\vec{\theta}) & = -E\left[ \frac{\partial^2}{\partial\mu \partial\beta} log~p(X; \vec{\theta}) \right] \\ \nonumber & = \frac{1}{\sigma^2}E\left\{ \Delta \left[ (x_i - \mu) - \beta\sum_{j\in\eta_i}(x_j - \mu) \right] + (1 - \beta\Delta)\left[ \sum_{j\in\eta_i}(x_j - \mu) \right] \right\} \\ \nonumber & = 0 + 0 = 0
\end{align}

Moving forward to the fourth component, note that $O = II_{\sigma^2\mu}(\vec{\theta}) = 0$, since a change in the order of the differentiation operators does not affect the result.  Thus, we proceed directly to the fifth component, given by:	

\begin{align}
	P = II_{\sigma^2\sigma^2} (\vec{\theta}) & = - E\left[ \frac{\partial^2}{\partial(\sigma^2)^2} log~p(X; \vec{\theta})  \right] \\ \nonumber & = - E \left\{ \frac{\partial}{\partial\sigma^2}\left[ -\frac{1}{2\sigma^2} + \frac{1}{2\sigma^4} \left( x_i - \mu -\beta\sum_{j\in\eta_i}(x_j - \mu) \right)^2 \right] \right\} \\ \nonumber & = - E \left\{ \frac{1}{2\sigma^4} - \frac{1}{\sigma^6}\left[ (x_i - \mu) - \beta\sum_{j\in\eta_i}(x_j - \mu) \right]^2 \right\} \\ \nonumber & = \frac{1}{2\sigma^4} - \frac{1}{\sigma^6}\left[ 2\beta\sum_{j\in\eta_i} \sigma_{ij} - \beta^2 \sum_{j\in\eta_i}\sum_{k\in\eta_i}\sigma_{jk} \right]
\end{align}

The next component of the second fundamental form is given by:

\begin{align}
	Q = II_{\sigma^2\beta} (\vec{\theta}) & = - E\left[ \frac{\partial^2}{\partial\sigma^2 \partial\beta} log~p(X; \vec{\theta})  \right] \\ \nonumber & = - E \left\{ \frac{\partial}{\partial\sigma^2}\left[ \frac{1}{\sigma^2} \left( x_i - \mu - \beta\sum_{j\in\eta_i}(x_j - \mu) \right)\left( \sum_{j\in\eta_i}(x_j - \mu) \right) \right]  \right\} \\ \nonumber & = \frac{1}{\sigma^4}\left[ \sum_{j\in\eta_i}\sigma_{ij} - \beta \sum_{j\in\eta_i}\sum_{k\in\eta_i}\sigma_{jk} \right]
\end{align} 

Note that the next two components of the second fundamental form are identical to their symmetric counterparts, that is, $R = II_{\beta\mu} (\vec{\theta}) = II_{\mu\beta} (\vec{\theta}) = N = 0$ and $S = II_{\beta\sigma^2}(\vec{\theta}) = II_{\sigma^2\beta}(\vec{\theta}) = Q$. Finally, the last component of the second fundamental form is given by:

\begin{align}
	T = II_{\beta\beta}(\vec{\theta}) = - E\left[ \frac{\partial^2}{\partial\beta^2} log~p(X; \vec{\theta}) \right]
\end{align} which can be computed as:

\begin{align}
	T & = -\frac{1}{\sigma^2}E \left\{\frac{\partial}{\partial\beta}\left[\left(  x_i - \mu - \beta\sum_{j\in\eta_i}(x_j - \mu)\right) \left( \sum_{j\in\eta_i}(x_j - \mu) \right) \right] \right\} \\ \nonumber & = \frac{1}{\sigma^2}E\left[ \left( \sum_{j\in\eta_i}(x_j - \mu) \right) \left( \sum_{j\in\eta_i}(x_j - \mu) \right) \right] \\ \nonumber & = \frac{1}{\sigma^2}\sum_{j\in\eta_i}\sum_{k\in\eta_i}\sigma_{jk}
\end{align} concluding that the second fundamental form has the following structure:

\begin{equation}
	II(\vec{\theta}) = \left( \begin{array}{ccc}
	L & 0 & 0 \\ 
	0 & P & Q \\ 
	0 & Q & T
	\end{array} \right)
\end{equation}

Finally, note also that when the inverse temperature parameter is fixed at zero, both first and second fundamental forms converge to:

\begin{equation}
	g(\vec{\theta}) = \left( \begin{array}{ccc}
	\frac{1}{\sigma^2} & 0 & 0 \\ 
	0 & \frac{1}{2\sigma^4} & 0 \\ 
	0 & 0 & 0
	\end{array} \right)
\end{equation} which is exactly the metric tensor (Fisher information matrix) of the parametric space of a single Gaussian random variable.

\subsection{Fundamental forms in tensorial notation}

In order to reduce the computational burden in the numerical computations, we propose to express the components of the first and second fundamental forms using Kronecker products (tensor products). First, note that we can convert each $3 \times 3$ neighborhood patch formed by $x_{i} \cup \eta_{i}$ into a vector $p_i$ of 9 elements by piling its rows. Then, we compute the covariance matrix of these vectors, for $i = 1, 2,..., n$ denoted by $\Sigma_{p}$. From this covariance matrix, we extract two main components: 1) a vector of size 8, $\vec{\rho}$, composed by the the elements of the central row of $\Sigma_{p}$, excluding the middle one, which denotes the variance of $x_i$ (we want only the covariances between $x_i$ and $x_j$, for $j \neq i$; and 2) a sub-matrix of dimensions $8 \times 8$, $\Sigma_{p}^{-}$, obtained by removing the central row and central column from $\Sigma_{p}$ (we want only the covariances between $x_j \in \eta_i$ and $x_k \in \eta_i$). Figure \ref{fig:cov_matrix} shows the decomposition of the covariance matrix $\Sigma_{p}$ into the sub-matrix $\Sigma_{p}^{-}$ and the vector $\vec{\rho}$. By employing Kronecker products, we rewrite the first fundamental form (metric tensor) in a tensorial notation, providing a computationally efficient way to compute the elements of $I(\vec{\theta})$:

\begin{figure}[ht]
\begin{center}
\includegraphics[scale=0.4]{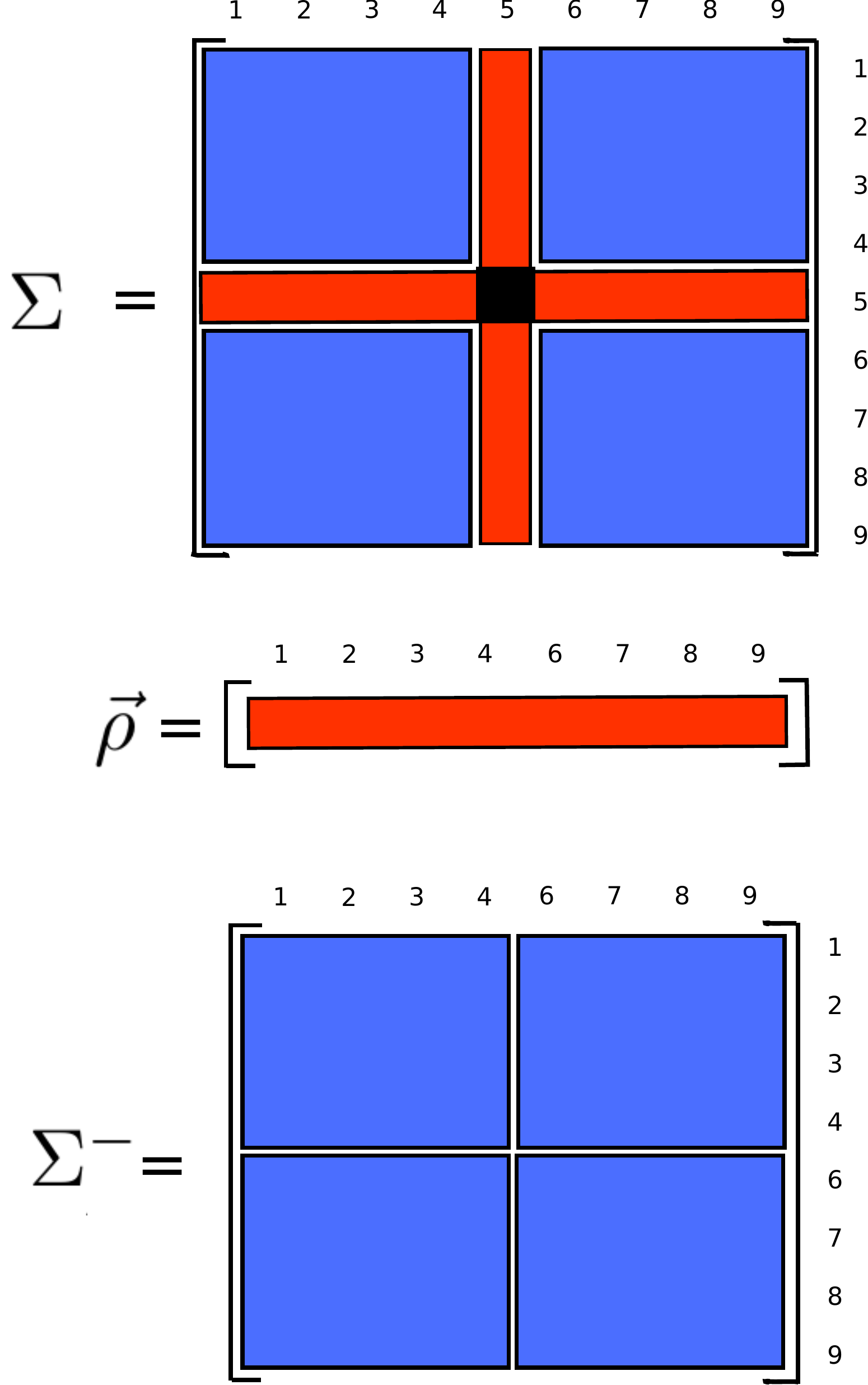}
\end{center}
\caption{Decomposition of $\Sigma_{p}$ into $\Sigma_{p}^{-}$ and $\vec{\rho}$ on a second-order neighborhood system ($\Delta=8$). By rewriting the components of the first and second fundamental forms in terms of Kronocker products, we can make numerical simulations faster.
}
\label{fig:cov_matrix}
\end{figure}

\begin{equation}
	A = I_{\mu\mu}(\vec{\theta}) = \frac{1}{\sigma^2}\left(1-\beta \Delta \right)^2\left[ 1 - \frac{1}{\sigma^2}\left( 2\beta\left\| \vec{\rho} \right\|_{+} - \beta^2 \left\| \Sigma_{p}^{-} \right\|_{+} \right) \right] 
\end{equation}

\begin{align}	
	E = I_{\sigma^2\sigma^2}(\vec{\theta}) = \frac{1}{2\sigma^4} & - \frac{1}{\sigma^6} \left[ 2\beta\left\| \vec{\rho} \right\|_{+} - \beta^2 \left\| \Sigma_{p}^{-} \right\|_{+} \right] \\ \nonumber & + \frac{1}{\sigma^8}\left[ 3\beta^2 \left\| \vec{\rho} \otimes \vec{\rho} \right\|_{+} - 3 \beta^3 \left\| \vec{\rho} \otimes \Sigma_{p}^{-} \right\|_{+} + 3\beta^4 \left\| \Sigma_{p}^{-} \otimes \Sigma_{p}^{-} \right\|_{+}  \right] \nonumber
\end{align}	
	
\begin{align}
	F = I_{\sigma^2\beta}(\vec{\theta}) = I_{\beta\sigma^2}(\vec{\theta}) & = \frac{1}{\sigma^4}\left[ \left\| \vec{\rho} \right\|_{+} - \beta \left\| \Sigma_{p}^{-} \right\|_{+} \right] \\ \nonumber & - \frac{1}{2\sigma^6} \left[ 6\beta \left\| \vec{\rho} \otimes \vec{\rho} \right\|_{+} - 9 \beta^2 \left\| \vec{\rho} \otimes \Sigma_{p}^{-} \right\|_{+} + 3\beta^3 \left\| \Sigma_{p}^{-} \otimes \Sigma_{p}^{-} \right\|_{+}  \right] \nonumber
\end{align}	

\begin{equation}	
	I = I_{\beta\beta}(\vec{\theta}) = \frac{1}{\sigma^2} \left\| \Sigma_{p}^{-} \right\|_{+} + \frac{1}{\sigma^4} \left[ 2 \left\| \vec{\rho} \otimes \vec{\rho} \right\|_{+} - 6 \beta \left\| \vec{\rho} \otimes \Sigma_{p}^{-} \right\|_{+} + 3\beta^2 \left\| \Sigma_{p}^{-} \otimes \Sigma_{p}^{-} \right\|_{+}  \right] 
\end{equation} where $\left\| A \right\|_{+}$ represents the summation of all the entries of the vector/matrix $A$ and $\otimes$ denotes the Kronecker (tensor) product. Similarly, by employing the same reasoning, the elements of the second fundamental form $II(\vec{\theta})$ can be expressed as:

\begin{equation}
	L = II_{\mu\mu}(\vec{\theta}) = \frac{1}{\sigma^2}\left(1-\beta \Delta \right)^2
\end{equation}

\begin{align}	
	P = II_{\sigma^2\sigma^2}(\vec{\theta}) = \frac{1}{2\sigma^4} - \frac{1}{\sigma^6} \left[ 2\beta\left\| \vec{\rho} \right\|_{+} - \beta^2 \left\| \Sigma_{p}^{-} \right\|_{+} \right]
\end{align}	
	
\begin{align}
	Q = II_{\sigma^2\beta}(\vec{\theta}) = II_{\beta\sigma^2}(\vec{\theta}) = \frac{1}{\sigma^4}\left[ \left\| \vec{\rho} \right\|_{+} - \beta \left\| \Sigma_{p}^{-} \right\|_{+} \right]
\end{align}	

\begin{equation}	
	T = II_{\beta\beta}(\vec{\theta}) = \frac{1}{\sigma^2} \left\| \Sigma_{p}^{-} \right\|_{+} 
\end{equation} 

\section{Geometric analysis of Gaussian random field dynamics}

\begin{figure}[!h]
	\begin{center}
	\includegraphics[scale=0.5]{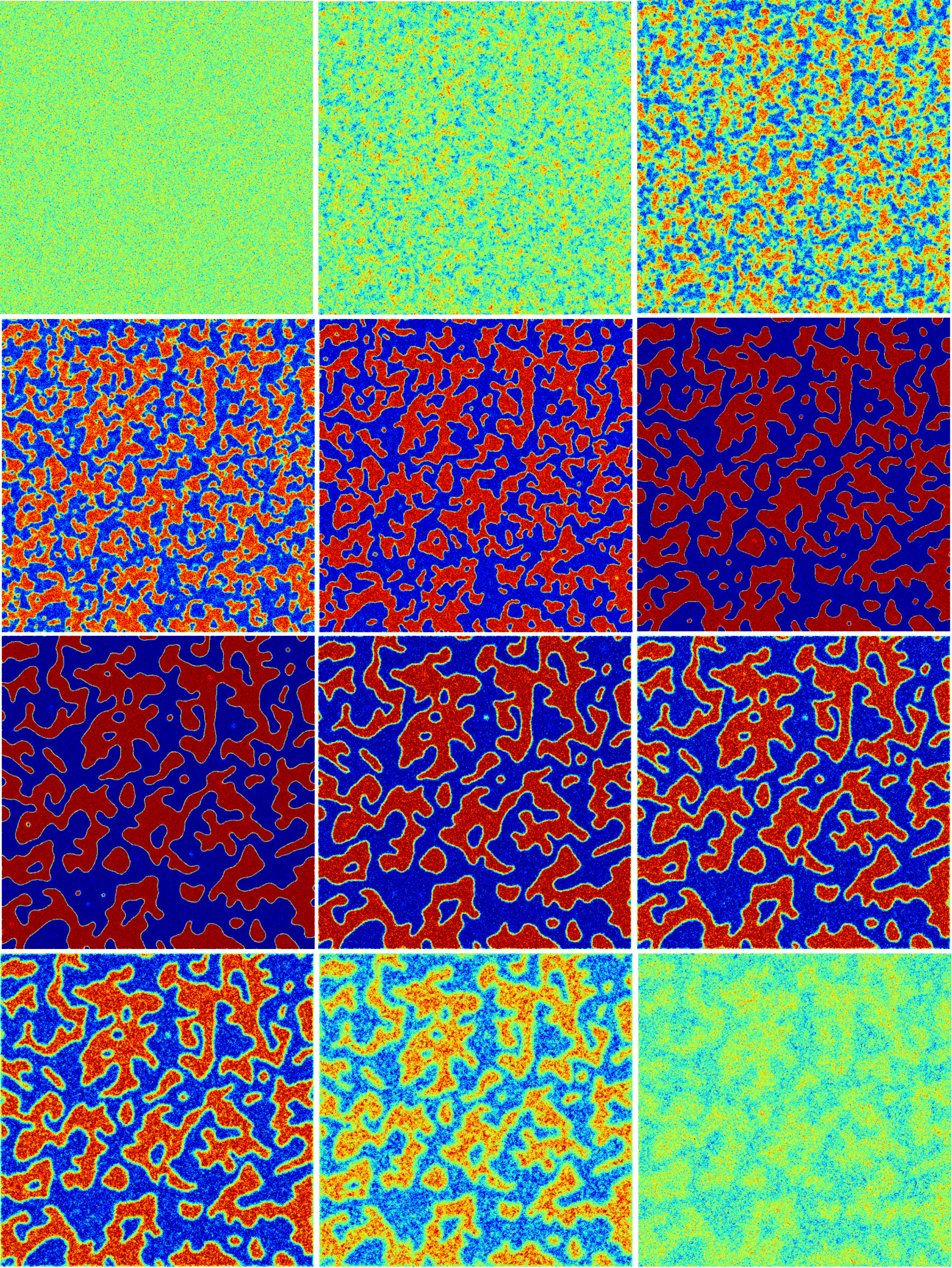}
	\end{center}
	\caption{Evolution of the random field as the inverse temperature parameter $\beta$ is first increased from zero to $\beta_{MAX}$ and then decreased to zero again.}
	\label{fig:MCMC}
\end{figure}

In order to simulate the dynamics of the Gaussian random field, Markov Chain Monte Carlo (MCMC) simulation was employed to generate outcomes of the model using the Metropolis-Hastings algorithm \cite{Metropolis}. For each simulation, arbitrary initial values for the parameters $\mu$ and $\sigma^2$ are selected, but the initial inverse temperature is set to zero, making all random variables independent. At the end of each iteration, we perform a small and positive small displacement $\Delta\beta$ until a maximum value $\beta_{MAX}$ is reached, defining the first half of an information cycle. After that, to complete a cycle, at each iteration, the same negative small displacement $-\Delta\beta$ is performed, until the inverse temperature reaches zero once again. The other parameters of the random field (mean and variance) are estimated by the sample mean and sample variance. In our simulations, a complete full cycle takes 1000 iterations, where each outcome is a $512 \times 512$ matrix. Figure \ref{fig:MCMC} shows some samples of the random field during the evolution of the system.

A relevant quantity in the study of stochastic complex systems is entropy, one of the most ubiquitous concepts in science, with applications in a large number of research fields. For instance, in information theory, entropy is related to the degree of uncertainty about a source of information \cite{shannon1949}. In statistical physics, entropy plays an important role, being a central piece in the second law of thermodynamics, which states that the entropy of isolated systems left to spontaneous evolution cannot decrease, as they always arrive at a state of thermodynamic equilibrium, where the entropy is highest \cite{SecondLaw}. 

To compute the entropy in a pairwise isotropic Gaussian-Markov random field, recall that it can be defined as the expected value of self-information, which leads to:

\begin{align}
	\label{eq:entropia1}
	 H_{\beta}(\vec{\theta}) & = - E\left[ log~p\left(x_{i}| \eta_{i}, \vec{\theta} \right) \right] \\ \nonumber & = \frac{1}{2}\left[ log\left( 2\pi\sigma^2 \right) + 1\right] - \frac{1}{\sigma^2} \left[ \beta\sum_{j \in \eta_i}\sigma_{ij} - \frac{\beta^2}{2}\sum_{j \in \eta_i}\sum_{k \in \eta_i}\sigma_{jk} \right] \\ \nonumber & = H_{G}(\vec{\theta}) - \frac{\beta}{\sigma^2}\sum_{j \in \eta_i}\sigma_{ij} + \frac{\beta^2}{2\sigma^2} \sum_{j \in \eta_i}\sum_{k \in \eta_i}\sigma_{jk}
\end{align} where $H_G(\vec{\theta})$ denotes the entropy of a Gaussian random variable. Note that the entropy is a quadratic function of the inverse temperature parameter $\beta$. Besides, for $\beta=0$, we have $H_{\beta}(\vec{\theta}) = H_{G}(\vec{\theta})$, as expected. Using the Kronecker product and rewriting the summations in terms of the components $Q$ and $T$ from the second fundamental form, we have:


\begin{equation}
	H_{\beta}(\vec{\theta}) = H_{G}(\vec{\theta}) - \beta \sigma^2 Q - \beta^2\frac{T}{2}
\end{equation}

\begin{figure}[!h]
	\begin{center}
	\includegraphics[scale=0.41]{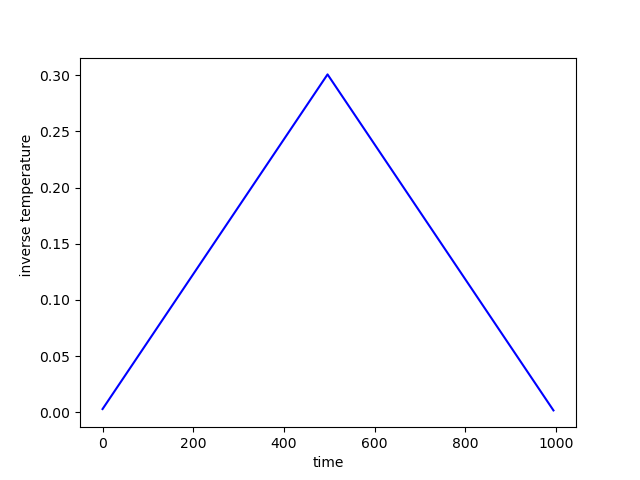}
	\includegraphics[scale=0.41]{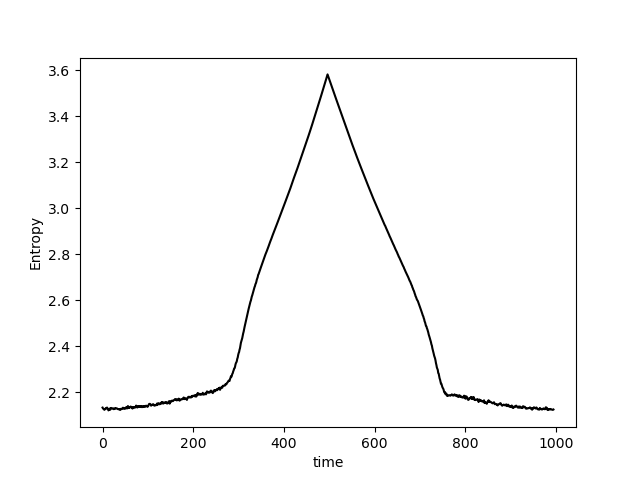}
	\end{center}
	\caption{Evolution of the inverse temperature parameter and entropy along a full cycle of the MCMC simulation with the Gaussian random field model.}
	\label{fig:Entropy}
\end{figure}

In order to understand the numerical simulations, it is necessary to explain how the results were obtained. First, a full cycle of the MCMC simulation is composed by 1000 iterations: initially, the inverse temperature parameter $\beta$ is set to zero, and at the end of each iteration, $\beta$ is incremented by $\Delta\beta = 0.0006$, up to the 500th iteration. In the second half of the cycle, at the end of each iteration, $\beta$ is decremented by $\Delta\beta = 0.0006$, until it reaches zero once again. Figure \ref{fig:Entropy} shows the variation of the inverse temperature and entropy along a full cycle.

\begin{figure}[!h]
	\begin{center}
	\includegraphics[scale=0.41]{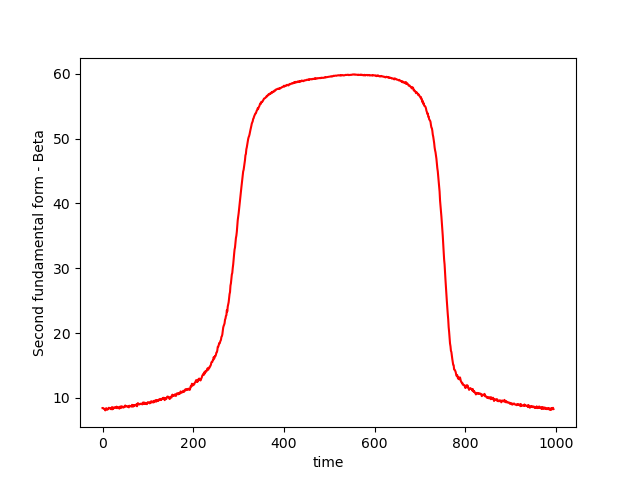}
	\includegraphics[scale=0.41]{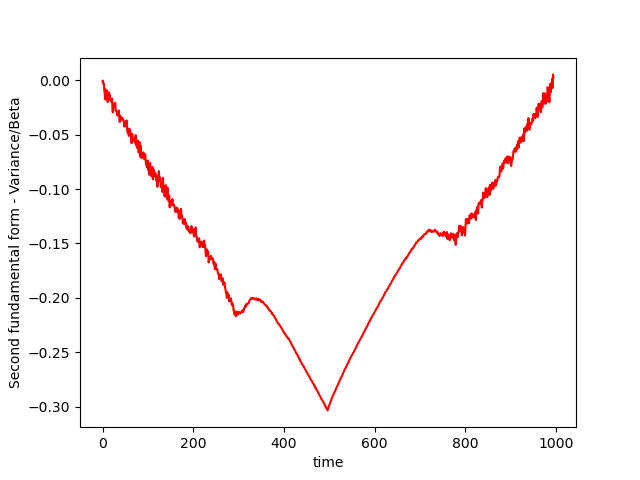}
	\end{center}
	\caption{Evolution of the components $T$ (left) and $Q$ (right) of the second fundamental form along a full cycle of the MCMC simulation with the Gaussian random field model.}
	\label{fig:second}
\end{figure}

As the entropy of a Gaussian random field depends directly on the components $Q$ and $T$ of the second fundamental form, we also plot the variations of these quantities along a full cycle of the MCMC simulation. Figure \ref{fig:second} illustrates how $Q$ and $T$ changes over time. Note that while the plot of $T$ is almost symmetric, the plot of $Q$ is highly asymmetric.

\begin{figure}[!h]
	\begin{center}
	\includegraphics[scale=0.68]{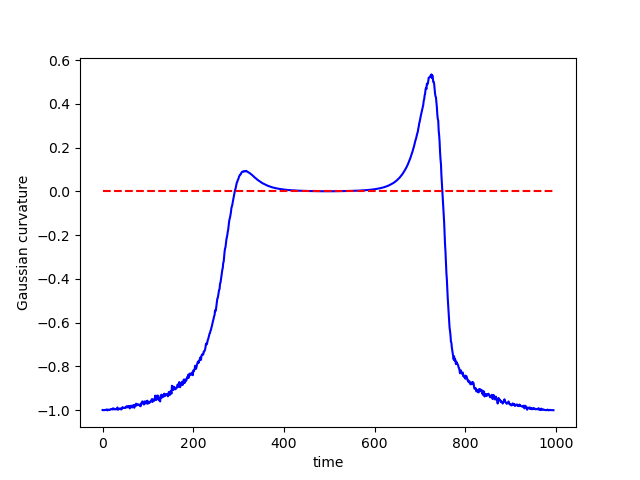}
	\end{center}
	\caption{Evolution of the Gaussian curvature along a full cycle of the MCMC simulation with the Gaussian random field model.}
	\label{fig:GC}
\end{figure}

The behavior of the Gaussian curvature along a full cycle of the numerical simulations shows an interesting pattern. In the beginning, when the inverse temperature is zero, the parametric space has constant negative Gaussian curvature ($K=-1$), which means hyperbolic geometry.  During the first half, when the system moves towards higher entropy states, the sign of the Gaussian curvature changes from negative to positive, whereas in the second half, when the system moves towards lower entropy states, the sign of the Gaussian curvature changes from positive to negative. However, the amount of curvature necessary to bend and stretch/shrink the parametric space when moving towards lower entropy states is significantly higher than that necessary to bend and stretch/shrink the parametric space when moving towards higher entropy states. We call this asymmetric pattern of evolution of Gaussian random fields as \emph{the curvature effect}, which can be described as: the variations of the Gaussian curvature when the system moves towards higher entropy states is different from the variations of the Gaussian curvature when the system moves towards lower entropy states. Figure \ref{fig:GC} illustrates the variation of the Gaussian curvature along the MCMC simulation.

\begin{figure}[!h]
	\begin{center}
	\includegraphics[scale=1]{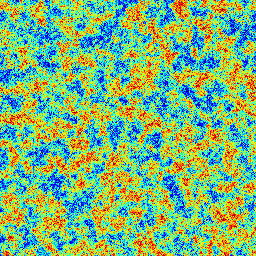}
	\includegraphics[scale=1]{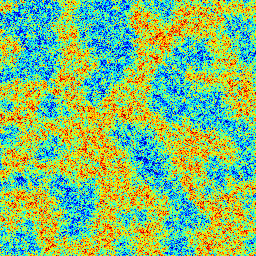}
	\end{center}
	\caption{Global configuration of a Gaussian random field when the Gaussian curvature change its sign. From left to right: a) Gaussian curvature becomes positive; b) Gaussian curvature becomes negative.}
	\label{fig:sinais}
\end{figure}

An interesting question about the results obtained with the MCMC simulation is: when exactly the Gaussian curvature becomes positive? Our numerical computations show that there is a change in the sign of the Gaussian curvature around $\beta = 0.178$ (297th iteration). And when exactly the Gaussian curvature becomes negative? Around $\beta = 0.148$ (755th iteration). These moments coincide with abrupt changes in the behavior of the system's entropy. Figure \ref{fig:sinais} shows the global configurations of the system in the points where the Gaussian curvature change its sign. It is possible to visualize that change in the sign of the Gaussian curvature is directly related to moments in which the random field is undergoing phase transitions. Therefore, our results suggest that the Gaussian curvature is a good measure to determine whether the system is close to a critical point, that is, if it is approaching a phase transition.

\begin{figure}[!h]
	\begin{center}
	\includegraphics[scale=0.41]{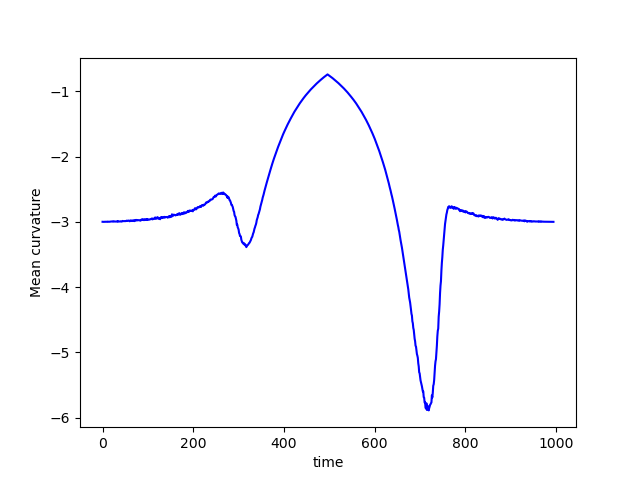}
	\includegraphics[scale=0.41]{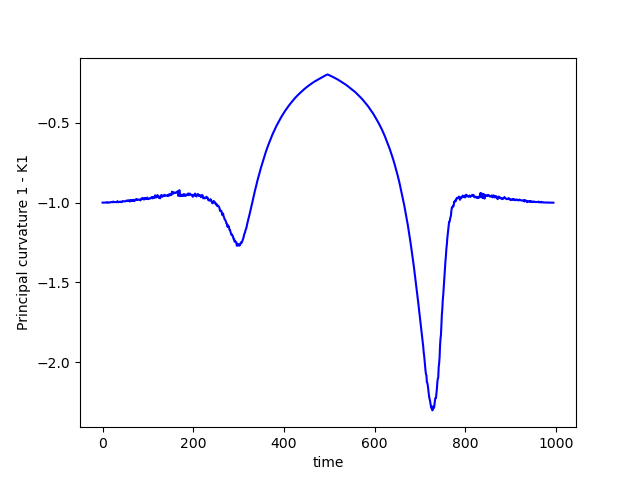}
	\includegraphics[scale=0.41]{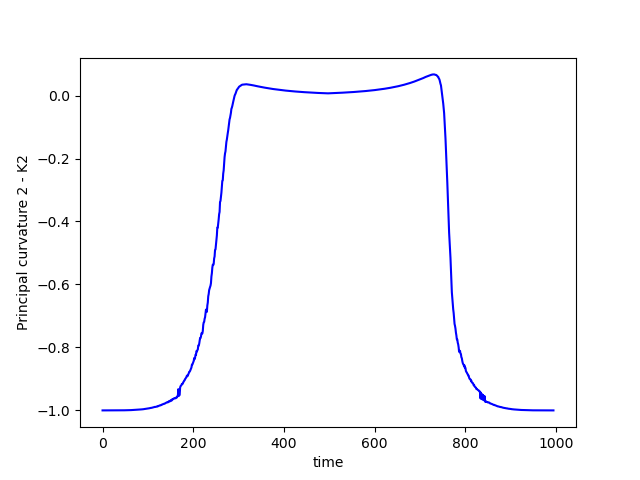}
	\includegraphics[scale=0.41]{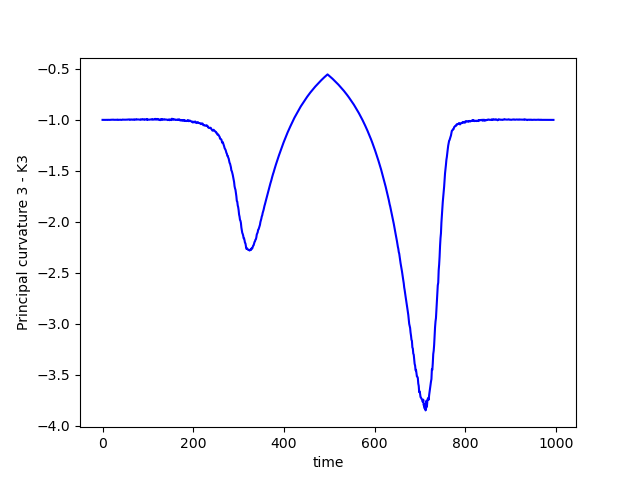}
	\end{center}
	\caption{Evolution of the mean and principal curvatures along a full cycle of the MCMC simulation with the Gaussian random field model.}
	\label{fig:MC}
\end{figure}

By inspecting the mean and the principal curvatures, we also note a highly asymmetric pattern of evolution. The difference in comparison with the Gaussian curvature is that the mean curvature is always negative due to the principal curvatures. The second principal curvature is the only one to become positive along the MCMC simulation. From differential geometry, we know that the mean curvature is the summation of the principal curvatures and the Gaussian curvature is the product of the principal curvatures. Figure \ref{fig:MC} illustrates the variation of the mean and principal curvatures along the MCMC simulation.

\begin{figure}[!h]
	\begin{center}
	\includegraphics[scale=0.41]{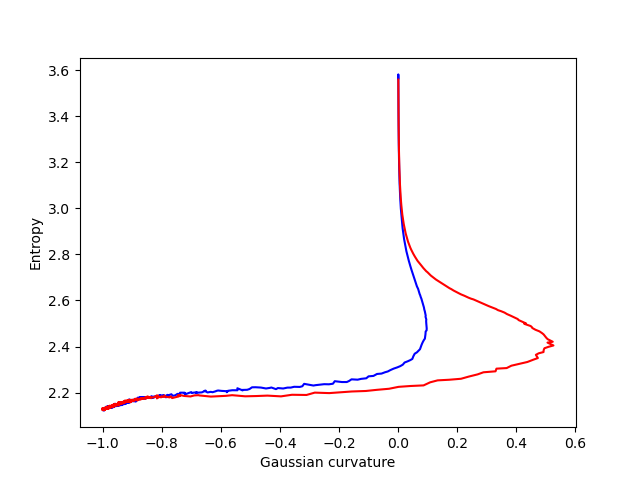}
	\includegraphics[scale=0.41]{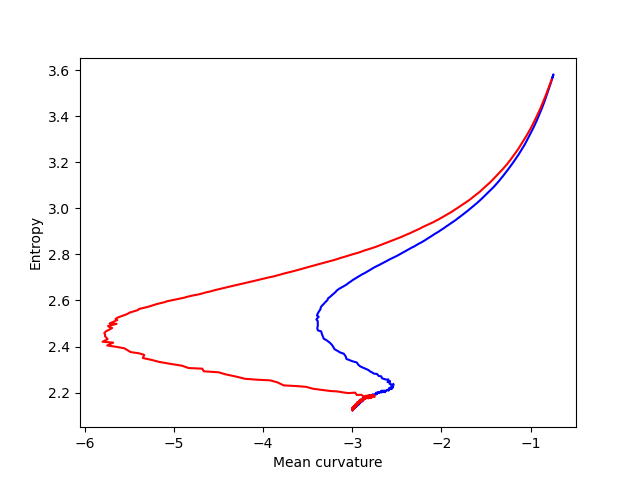}
	\end{center}
	\caption{The amount of stretching/shrinking the parametric space of Gaussian random fields suffers when it bends depends whether the entropy is increasing (blue curve) or decreasing (red curve). From left to right: a) Gaussian curvature; b) Mean curvature.}
	\label{fig:curvature_path}
\end{figure}

In order to analyze how the system's entropy change as a function of the curvatures, we build a 2D representation to visualize the entropy as a function of the curvatures. The idea is to visualize what happens to the entropy as curvature changes along the MCMC simulation. In fact, what we observe is that entropy is not a function of the curvature, since for some values of curvature we have two different values of entropy. This \emph{curvature effect} indicates that the amount of curvature needed to bend the space during an increase in the system's entropy (blue line) is smaller than the amount of curvature required to bend the space during the same reduction in the system's entropy (red curve). Figure \ref{fig:curvature_path} shows the curvature paths for both Gaussian and mean curvatures.

Note that this behavior induces a natural orientation to the process of, from a low entropy state, bringing the random field to a high entropy state and back, which is an intrinsic notion of time, as the variation of curvature in the parametric space (according to the \emph{Theorema Egregium}, the Gaussian curvature does not depend on an extrinsic referential). It is interesting to observe that in the Gaussian curvature cycle (left image), the natural orientation is clockwise (blue curve first, red curve after), while in the mean curvature cycle (right image), the natural orientation is anti-clockwise (blue curve first, red curve after). The Python source code with the computational implementation of the numerical simulations are available at: \url{https://github.com/alexandrelevada/Curvature_GMRF}.

\section{Conclusions}
Stochastic systems are composed of several random variables that interact in a non-linear way along time. Depending on how the parameter models change, a series of complex behavior can emerge from this dynamics. In this study, we addressed the problem of deriving closed-form expressions for the first and second fundamental forms of the underlying manifolds of Gaussian random fields in order to compute the principal, mean and Gaussian curvatures. Information geometry shows that the fundamental forms of parametric spaces of random variables are given by the first and second-order Fisher information matrices of the statistical model. Mathematical expressions for the components of these matrices were derived, allowing the computation of intrinsic geometric properties.  

The intrinsic geometric structure of statistical models of independent random variables has been extensively studied in information geometry. For instance, it has been shown that the parametric space of a Gaussian random variable has constant negative Gaussian curvature, which means that the geometry is hyperbolic. However, little is known about the geometry of random field models, where the inverse temperature parameter induces a spatial dependence structure among the variables.  In this paper, we investigated how the variation of the system's entropy is related to the variation in the principal, mean and Gaussian curvatures. 

Using MCMC simulations via the Metropolis-Hastings algorithm, we performed several cycles of evolution composed by a first stage in which the system's entropy is increased, and a second stage in which the system's entropy is decreased to the minimum value. The obtained results show that the variations in principal, mean and Gaussian curvatures are highly asymmetric, suggesting that the parametric space suffers a series of irreversible geometric deformations. Our geometric analysis has shown an unreported phenomenon: the \emph{curvature effect}, which suggests that the deformations in the parametric space are more prominent during a decrease of the inverse temperature than during an increase of the inverse temperature, indicating the emergence of an arrow of time in the evolution of the random field.


Future works may include a deeper study about the relationship between curvature and the geodesic distances between two random fields operating in different regimes, as a way to provide an intrinsic similarity measure. An analysis on how the components of the first and second fundamental form change as the inverse temperature varies may reveal relevant information about the underlying geometric structure of the parametric space. Besides, we intend to investigate techniques for the estimation of the inverse temperature parameter in order to simulate a situation in which we do not have direct access to the real inverse temperature value. Another idea consists in computing information-theoretic divergences, such as the KL-divergence, between pairs of random field models. Furthermore, we intend to study the feasibility of the application of the proposed information-geometric analysis in other random field models, such as the classic binary Ising model and the q-state Potts models, in which each variable assumes one of $q$ different discrete states. Finally, we intend to compute approximations for the geodesic distances between Gaussian-Markov random fields and develop machine learning applications to perform unsupervised metric learning in classification problems.

\section*{Acknowledgements}
This study was financed in part by the Coordenação de Aperfeiçoamento de Pessoal de Nível Superior - Brasil (CAPES) - Finance Code 001.


\bibliography{mybibfile}

\begin{thebibliography}{10}
\expandafter\ifx\csname url\endcsname\relax
  \def\url#1{\texttt{#1}}\fi
\expandafter\ifx\csname urlprefix\endcsname\relax\def\urlprefix{URL }\fi
\expandafter\ifx\csname href\endcsname\relax
  \def\href#1#2{#2} \def\path#1{#1}\fi

\bibitem{Paolo}
P.~Sibani, H.~J. Jensen, Stochastic Dynamics of Complex Systems: From Glasses
  to Evolution, Imperial College Press, 2013.

\bibitem{Baryam}
Y.~Bar-yam, Dynamics of Complex Systems, CRC Press, 1999.

\bibitem{Automata}
B.~Chopard, M.~Droz, Cellular Automata Modeling of Physical Systems, Cambridge
  University Press, 2005.

\bibitem{ComplexNetworks}
R.~Albert, A.~L. Barab\'asi, Statistical mechanics of complex networks, Rev.
  Mod. Phys. 74 (2002) 47--97.

\bibitem{RandomFields}
A.~Bulinski, E.~Spodarev, Introduction to Random Fields, Springer Berlin
  Heidelberg, Berlin, Heidelberg, 2013, pp. 277--335.

\bibitem{Willsky}
A.~Willsky, N.~Sandell, The stochastic analysis of dynamic systems moving
  through random fields, IEEE Transactions on Automatic Control 27~(4) (1982)
  830--838.

\bibitem{Phase}
C.~Merkle, C.~Rohde, Computation of dynamical phase transitions in solids,
  Applied Numerical Mathematics 56~(10) (2006) 1450--1463.

\bibitem{StatMech}
S.~Friedli, Y.~Velenik, Statistical Mechanics of Lattice Systems: A Concrete
  Mathematical Introduction, Cambridge University Press, 2017.

\bibitem{Gibbs}
J.~W. Gibbs, Elementary Principles in Statistical Mechanics, Charles Scribner's
  Sons, 1902.

\bibitem{MRFBiol}
W.~G.W., E.~S.R., Remote homology search with hidden potts models, PLoS
  Computational Biology 16~(11) (2020) 1--22.

\bibitem{MRFEcon}
L.~Onural, M.~Pınar, C.~Fırtına, Modeling economic activities and random
  catastrophic failures of financial networks via gibbs random fields,
  Computational Economics 58 (2021) 203--232.

\bibitem{Ising}
B.~A. Cipra, An introduction to the ising model, The American Mathematical
  Monthly 94~(10) (1987) 937--959.

\bibitem{Potts}
F.~Y. Wu, The potts model, Rev. Mod. Phys. 54 (1982) 235--268.

\bibitem{GaussianRandomFields}
D.~T. Hristopulos, Gaussian Random Fields, Springer Netherlands, Dordrecht,
  2020, pp. 245--307.

\bibitem{MRFApp}
R.~Kindermann, J.~L. Snell, Markov Random Fields and their Applications,
  American Mathematical Society, 1980.

\bibitem{Wheeler}
J.~A. Wheeler, On the nature of quantum geometrodynamics, Annals of Physics
  2~(6) (1957) 604--614.

\bibitem{Geom2014}
M.~A. Scheel, K.~S. Thorne, Geometrodynamics: the nonlinear dynamics of curved
  spacetime, Physics-Uspekhi 57~(4) (2014) 342--351.

\bibitem{Unification}
I.~Ita, Eyo~Eyo, C.~Soo, H.-L. Yu, Intrinsic time quantum geometrodynamics,
  Progress of Theoretical and Experimental Physics 2015~(8) (2015) 083E01.

\bibitem{Nature}
K.~Y. Bliokh, A.~Niv, V.~Kleiner, E.~Hasman, Geometrodynamics of spinning
  light, Nature Photon 2 (2015) 748--753.

\bibitem{Amari}
S.~ichi Amari, H.~Nagaoka, Methods of Information Geometry, American
  Mathematical Society, 2000.

\bibitem{Frieden2004}
B.~R. Frieden, Science from Fisher Information, Cambridge University Press,
  2004.

\bibitem{Dodson}
K.~Arwini, C.~T.~J. Dodson, Information Geometry: Near Randomness and Near
  Independence, Springer, 2008.

\bibitem{Pinele}
J.~Pinele, J.~E. Strapasson, S.~I.~R. Costa, The fisher–rao distance between
  multivariate normal distributions: Special cases, boundsand applications,
  Entropy 22~(4) (2020) 404.

\bibitem{RFields}
E.~Hernández-Lemus, Random fields in physics, biology and data science,
  Frontiers in Physics 9 (2021) 77.

\bibitem{GRFs}
D.~T. Hristopulos, Random Fields for Spatial Data Modeling: A Primer for
  Scientists and Engineers, Springer, 2020, Ch. Gaussian Random Fields, pp.
  245--307.

\bibitem{CLT}
B.~Davis, D.~McDonald, An elementary proof of the local central limit theorem,
  Journal of Theoretical Probability 8 (1995) 693--701.

\bibitem{MRF_DNN}
A.~Arnab, S.~Zheng, S.~Jayasumana, B.~Romera-Paredes, M.~Larsson, A.~Kirillov,
  B.~Savchynskyy, C.~Rother, F.~Kahl, P.~H. Torr, Conditional random fields
  meet deep neural networks for semantic segmentation: Combining probabilistic
  graphical models with deep learning for structured prediction, IEEE Signal
  Processing Magazine 35~(1) (2018) 37--52.

\bibitem{Review_Ising}
N.~G. Fytas, V.~Martín-Mayor, M.~Picco, N.~Sourlas, Review of recent
  developments in the random-field ising model, Journal of Statistical Physics
  172 (2018) 665--672.

\bibitem{Peacock}
J.~A. Peacock, Cosmological Physics, Cambridge University Press, 1998.

\bibitem{Hammersley}
J.~M. Hammersley, P.~Clifford,
  \href{www.statslab.cam.ac.uk/~grg/books/hammfest/hamm-cliff.pdf}{Markov field
  on finite graphs and lattices (preprint)} (1971).
\newline\urlprefix\url{www.statslab.cam.ac.uk/~grg/books/hammfest/hamm-cliff.pdf}

\bibitem{SeanCarroll}
S.~Carroll, Spacetime and Geometry: An Introduction to General Relativity,
  Addison-Wesley Professional, 2003.

\bibitem{Manfredo}
M.~P. do~Carmo, Differential Geometry of Curves and Surfaces, 2nd Edition,
  Dover Publications Inc., 2017.

\bibitem{Bar}
C.~Bär, Elementary Differential Geometry, Cambridge University Press, 2010.

\bibitem{Oneill}
B.~O'Neill, Elementary Differential Geometry, 2nd Edition, Elsevier, 2006.

\bibitem{Pressley}
A.~Pressley, Elementary Differential Geometry, 2nd Edition, Springer, 2012.

\bibitem{Woodward}
L.~Woodward, J.~Bolton, A First Course in Differential Geometry: Surfaces in
  Euclidean Space, Cambridge University Press, 2019.

\bibitem{Nielsen}
F.~Nielsen, An elementary introduction to information geometry, Entropy 22~(10)
  (2020) 1--61.

\bibitem{Quantum}
K.~T. Grosvenor, Information geometry and quantum fields, in: F.~Barbaresco,
  F.~Nielsen (Eds.), Geometric Structures of Statistical Physics, Information
  Geometry, and Learning, Springer International Publishing, 2021, pp.
  330--341.

\bibitem{Bickel}
P.~J. Bickel, Mathematical Statistics, Holden Day, New York, NY, USA, 1991.

\bibitem{Lehman}
E.~L. Lehmann, G.~Casella, Theory of Point Estimation, 2nd Edition,
  Springer-Verlag, New York, NY, USA, 1998.

\bibitem{Casella}
G.~Casella, R.~Berger, Statistical Inference, {Duxbury Resource Center}, 2001.

\bibitem{Silvey}
S.~D. Silvey, Statistical Inference, Chapman and Hall/CRC, 1975.

\bibitem{isserlis1918}
L.~Isserlis, On a formula for the product-moment coefficient of any order of a
  normal frequency distribution in any number of variables, Biometrika 12
  (1918) 134--139.

\bibitem{Metropolis}
N.~Metropolis, A.~W. Rosenbluth, M.~N. Rosenbluth, A.~H. Teller, E.~Teller,
  Equation of state calculations by fast computing machines, The Journal of
  Chemical Physics 21~(6) (1953) 1087--1092.

\bibitem{shannon1949}
C.~Shannon, W.~Weaver, The Mathematical Theory of Communication, University of
  Illinois Press, 1949.

\bibitem{SecondLaw}
R.~Jaffe, W.~Taylor, The Physics of Energy, Cambridge University Press, 2018.

\end{thebibliography}

\end{document}